\documentclass[a4paper,11pt]{article}
\pdfoutput=1 % if your are submitting a pdflatex (i.e. if you have
\usepackage{jcappub}
\usepackage{morefloats}

\newcommand{\be}{\begin{equation}}
\newcommand{\ee}{\end{equation}}
\newcommand{\beq}{\begin{eqnarray}}
\newcommand{\eeq}{\end{eqnarray}}

\def \lta {\mathrel{\vcenter{\hbox{$<$}\nointerlineskip\hbox{$\sim$}}}}

\def\t13{\mathrel{{\theta_{13}}}}
\def\y12{\mathrel{{\tan^2 \theta_{12}}}}
\def\c2{\mathrel{{\chi^2 }}}

\def \deg		{$^{\circ}$}
\def \sun	 	{$_{\odot}$}

\newcommand{\n}{neutrino}
\newcommand{\ns}{neutrinos}
\newcommand{\ic}{IceCube}

\newcommand{\mymail}{Kimberly.Emig@asu.edu}

\title{Do high energy astrophysical neutrinos trace star formation?}

\author[a]{Kimberly Emig,}
\author[b]{Cecilia Lunardini,}
\author[a]{and Rogier Windhorst}

\affiliation[a]{School of Earth and Space Exploration, Arizona State University, \\ 450 E. Tyler Mall, Tempe, AZ 85287-1404 USA}
\affiliation[b]{Department of Physics, Arizona State University, \\ 450 E. Tyler Mall, Tempe, AZ 85287-1504 USA}
\emailAdd{\mymail}
\emailAdd{Cecilia.Lunardini@asu.edu}
\emailAdd{Rogier.Windhorst@asu.edu}

\abstract{
The IceCube Neutrino Observatory has provided the first map of the high energy ($\sim$ 0.01 -- 1 PeV) sky in neutrinos. Since neutrinos propagate undeflected, their arrival direction is an important identifier for sources of high energy particle acceleration. Reconstructed arrival directions are consistent with an extragalactic origin, with possibly a galactic component, of the neutrino flux. We present a statistical analysis of positional coincidences of the IceCube neutrinos with known astrophysical objects from several catalogs. When considering starburst galaxies with the highest flux in gamma-rays and infrared radiation, up to $n=8$ coincidences are found, representing an excess over the $\sim4$ predicted for the randomized, or ``null" distribution. The probability that this excess is realized in the null case, the $p$-value, is  $p=0.042$. This value falls to $p=0.003$ for a partial subset of gamma-ray-detected starburst galaxies and superbubble regions in the galactic neighborhood. Therefore, it is possible that starburst galaxies, and the typically hundreds of superbubble regions within them, might account for a portion of IceCube neutrinos. The physical plausibility of such correlation is discussed briefly. 
}


\arxivnumber{1507.05711}
      \subheader{\hfill INT-PUB-15-030}

\begin{document}
\maketitle
\flushbottom

\section{Introduction} 
\label{sec:intro}

Extra-solar \n\ astronomy is an infant science, 
born in 2013, with the first detection of astrophysical \ns\ of energies up to $\sim$PeV at the \ic\ experiment in Antarctica 
\cite{Aartsen2013b, Aartsen2013, Aartsen2014a}. The origin of these \ns\ has not been established yet and represents 
an important goal to learn about 
the fundamental physics at play in astrophysical accelerators.  
The inherent properties of neutrinos -- neutral, weakly interacting -- offer unique probes into relatively unknown high energy mechanisms such as stellar core collapse and jet formation, particle acceleration in magnetic fields, shockwave propagation, etc.

Theoretically, there is a close relationship between \ns\  and cosmic ray protons (CRp). Comparable fluxes of \ns\ and gamma rays are expected as by-products of CRp interactions.  Neutrinos can be created in proton-proton ($pp$) 
interactions and subsequent cascades of charged and neutral pions:
\begin{flalign}
p + p \rightarrow& \,\, \pi^0 + \pi^{\pm} + anything & \nonumber \\
\pi^0 \rightarrow & \,\, \gamma + \gamma \nonumber  \\
\pi^{\pm} \rightarrow& \,\, \mu^{\pm} + \nu_{\mu}(\bar{\nu}_{\mu}) \nonumber \\
\mu^{\pm} \rightarrow& \,\, e^{\pm} + \nu_e (\bar{\nu}_e) + \bar{\nu}_{\mu} (\nu_{\mu})
\label{ppchain}
\end{flalign}
or in proton-photon ($p\gamma$) interactions, e.g., with cosmic microwave background (CMB) radiation:
\begin{flalign}
p + \gamma \rightarrow& \, n + \pi^{+} / \, p + \pi^0 &
\end{flalign}
with subsequent pion decays as in eq. (\ref{ppchain}).
Neutrinos created in the above interactions have $\sim 5 \%$ of the initial CRp energy and 
$\sim 50-75 \%$ of the gamma-ray energy \citep{Aartsen2014a}.

In light of the \n-gamma-ray connection, the search for sources of the \ic\ \ns\ has turned to the most powerful known gamma-ray emitters, and, more broadly, to the objects that show high energy activity.
Since the \ic\ data are consistent with a diffuse flux, several authors have compared their  energy distribution  with the spectra predicted theoretically for the diffuse \n\ flux for several classes of possible sources.  It was found that the $pp$ mechanism appears to naturally fit the data (see e.g., \cite{Murase:2014tsa}), and, among the  $pp$-dominated objects, starburst and star-forming galaxies have emerged as a particularly interesting possibility, fitting the data well both in spectrum and normalization \cite{Murase2013,Liu2014,Tamborra2014,Chang2015}.  In these objects, both the elements required for abundant \n\ production -- proton acceleration and a dense proton background -- are expected. Indeed, starburst and star-forming galaxies are characterized by their high rate of star formation, which implies a high rate of proton-accelerating jets in core collapse supernovae and/or supernova remnants. Because star formation typically occurs in dense hydrogen clouds, these galaxies should also be good proton absorbers, and thus efficient \n\ sources \cite{Loeb:2006tw,Stecker:2006vz,Thompson:2006np}.  

In parallel, searches have been conducted for positional associations of the \n\ data to specific objects. This task is made difficult by the poor angular resolution of \ic\ ($\sim$15\deg\ \cite{Aartsen2013}). However, positional matching is attractive because it is practically model-independent, relying only on the fact that -- in the absence of exotic physics -- \ns\ propagate undeflected from the production point to Earth. 
 The searches performed by the \ic\ collaboration, including point-like and extended-emission sources, all gave results consistent with background only \citep{Aartsen2014a, Aartsen2014b, IceCubeCollaboration2014a}.  Other authors have pointed out non-significant associations of some of the data with galactic objects, mainly the Galactic Center \citep{Razzaque2013,Bai2014} and $Fermi$ bubbles \citep{Razzaque2013,Lunardini2014}. Coincidences with up to 93\% confidence level have also been noted with the arrival directions of ultra-high energy cosmic rays (UHECR) \cite{Fang2014,Moharana2015}.  Several classes of extragalactic point-sources have been examined as well for spatial associations, in particular blazars \cite{Padovani2014, Sahu2014, Krauss:2014tna, Krauss:2015pja, Petropoulou2015, ANTARES2015,Brown2015, Glusenkamp2015}, Seyfert galaxies \cite{Moharana2015} and star-forming galaxies  with high luminosities in hydrogen cyanide   
(HCN)  emission \cite{Anchordoqui2014}.  The conclusions were mixed in these cases. 

At this time, the status of searches for positional associations of the \ic\ data with astrophysical objects is still heterogeneous, and no consistent picture has emerged. 
There remains a need for advancement towards a more systematic and interdisciplinary approach, that can fully incorporate knowledge and techniques from both \n\ physics and astronomy.  This approach could be used to test the hypothesis that high energy \ns\ ultimately originate from star formation. 

This paper is meant to be a step in this direction. We perform a statistical analysis of the \ic\ neutrino data, to test for spatial coincidences with the brightest candidate sources from several catalogs.
Compared to previous literature, our work has a stronger focus on star formation as a possible origin of the \n\ events. It is on this subject that our results are the most significant. \\

Section~\ref{sec:data} proceeds with a description of the \ic\ \n\ data set and the catalogs that we have considered for counterparts. 
In section~\ref{sec:method}, we discuss the statistical method used.
The motivation and results of the statistical analyses focusing on star-forming activity are presented in 
section~\ref{sec:results}, followed by a
discussion of the possible role of starburst galaxies and nearby star-forming regions in section~\ref{sec:discuss}.
Additional testing of candidate sources for which we found null results has been included in appendix~\ref{appB}.
Conclusions are given in section~\ref{sec:conclude}. Where relevant, we use the Planck 2015 cosmological parameters ($H_0 = 68$ km s$^{-1}$ Mpc$^{-1}$, $\Omega_M = 0.32$, $\Omega_{\Lambda} = 0.68$) \cite{Planck2015}.

%%%%%%%%%%%%%%%%%%%%%%%%%%%%%%%%

\section{Data}
\label{sec:data}

\subsection{IceCube neutrino detections}
\label{sec:iceevents}

We consider the 37 data events obtained by \ic\  after 988 days of running \cite{Aartsen2014a}. When needed, individual events will be referred to by their number as in Table 1 of ref. \cite{Aartsen2014a}.  Each event is classified as track-like or shower-like, depending on its topology in the detector.
The track-like events occur when a neutrino interaction results in a particle 
shower with a discernible muon track and therefore have smaller angular resolution on the sky ($<1$\deg). 
The nine observed track-like events are consistent with the expected background of $8.4 \pm 4.2$ atmospheric muons.
Shower-like events, 
on the other hand, result in a spherical light pattern produced by particle showers with no discernible muon and therefore have poorer angular resolution (median $\sim$15\deg; 50\% confidence level of positional errors). The twenty-eight shower-like events are in excess of the expected background of $6.6^{+5.9}_{-1.6}$ atmospheric neutrinos.
Events 28 and 32 have coincident detections at the Ice Top surface array. Thus, they have been identified as background \cite{Aartsen2014a}
and will be excluded from our analysis. 

The 35 events we use are shown in e.g., figure~\ref{fig:bz}, in equatorial J2000 coordinates, with their median angular errors. Because \ic\ is located near the South Pole, its horizon coincides with the celestial Equator. Due to absorption of \ns\ in the Earth, the detector is considerably less sensitive to up-going neutrinos (i.e. below the horizon, as coming from the northern sky)
 compared with down-going neutrinos (i.e. above the horizon, from the
southern sky). The difference in performance increases with increasing \n\ energy --- at $E \sim {\rm PeV}$, the Earth is 
essentially opaque. 
This feature explains the noticeable asymmetry in the event distribution between the two hemispheres as seen in e.g., figure~\ref{fig:bz}. 

\subsection{Catalogs of possible counterparts}
\label{sec:cat}

When searching astronomical catalogs for possible counterparts of the \ic\ \ns, it is logical to consider 
gamma-ray emitters detected in the same energy range ($E \gtrsim 100$ TeV) 
as the \ic\ events. 
However, gamma-rays with $E_{\gamma} \gtrsim 100$ GeV can suffer from absorption due to photon interactions with 
extragalactic background light and at $E_{\gamma} \gtrsim$ PeV, interactions with the cosmic microwave background \cite{Hauser2001}. Furthermore, current TeV observations lack
uniform and complete all-sky coverage, which is one of the conditions of validity of our analysis 
(see section \ref{sec:method}). Therefore, 
we resort to  primarily using observations from the $Fermi$ Large Area Telescope (hereafter, $Fermi$-LAT) \cite{Atwood:2009ez} with sensitivity up to $\sim 300$ GeV, and specifically the $Fermi$-LAT 3FGL catalog \cite{FermiLATCollaboration2014}. 
We then supplement the analysis 
with TeVCat detections and a catalog of starburst galaxies based on their infra-red (IR) flux \cite{Becker2009, Sanders2003}. 

Considering that many modern, all-sky catalogs include thousands of objects, it was necessary to apply selection criteria to each
catalog. 
Two main principles are used to choose sensible selection criteria. The first is uniformity: each set of candidates is made of sources of the same type/morphology (e.g., blazars, Seyfert galaxies, etc.).  The second principle is the assumption that, in a given class of viable candidates, those that appear the brightest in photon flux should also be the brightest in \n\ flux, and therefore most likely to be responsible for the \n\ events. Hence, lower limits of photon flux at appropriate wavelengths will be imposed. 

Finally, we emphasize that our selection procedure is completely blind with respect to the position of a neutrino source candidate in the sky.

We investigated the following catalogs. 

\paragraph{TeVCat.} TeVCat\footnote{http://tevcat.uchicago.edu/} is a compilation of currently and previously known TeV gamma-ray sources. 
TeV gamma-ray instruments, sensitive to energies between 100 GeV -- 100 TeV, 
can image gamma-ray emission via atmospheric Cherenkov telescope arrays. 
The procedure of reconstructing particle showers, created from the interaction of gamma-rays with the Earth's atmosphere, 
allows an angular resolution of photon arrival typically of $<$ 0.1\deg~at 1 TeV \cite{Hinton2009, Holder2012}. 

\paragraph{3FGL.} \textit{Fermi}-LAT is a pair-production gamma-ray instrument operating in the 
20 MeV -- 300 GeV range, and the current, main workhorse of space based gamma-ray observations. 
Its angular resolution varies from $\sim$5\deg~at 100 MeV to 0.8\deg~at 1 GeV.
The LAT 4-year Point Source Catalog (3FGL) covers the entire sky for at least 15 Ms of observing time,
reaching a detection threshold of $ \simeq 3 \times 10^{-12}$ erg cm$^{-2}$ s$^{-1}$ in the 100 MeV --- 100 GeV energy range \cite{FermiLATCollaboration2014}. 
It is known that at low galactic latitudes ($|b| < 10$\deg) diffuse emission from pion decay, bremsstrahlung, and inverse Compton
scattering reduces the sensitivity, although it is hard to quantify by exactly how much \cite{FermiLATCollaboration2014}. 

\paragraph{IRAS.} We use flux measurements in the mid to far IR as an indicator of star-forming activity in starburst galaxies 
\cite{Becker2009}. In particular, we chose objects with the highest fluxes at 100 $\mu$m, which is 
close to the spectral luminosity peak of heated dust in starburst galaxies. The 100 $\mu$m measurements were all 
gathered with the Infrared Astronomical Satellite (IRAS) Revised Bright Galaxy Sample \cite{Sanders2003}, a sample of galaxies
chosen to have 60 $\mu$m flux density greater than 5.24 Jy, \footnote{1 Jy (Jansky)$= 10^{-23}$ erg s$^{-1}$ cm$^{-2}$ Hz$^{-1}$.} which covers 93\% of the sky excluding only a strip within 5\deg\ of the galactic plane ($|b| < 5$\deg).

%%%%%%%%%%%%%%%%%%%%%%%%%%%%%%%%

\section{Statistical method}
\label{sec:method}

We adopt a version of the likelihood ratio statistical method which is commonly used in astronomy (see, e.g. \cite{deRuiter1977, Windhorst1984, Sutherland1992}) 
and has been used in high energy astrophysics \cite{Virmani2002,Moharana2015}
to test the spatial correlation between a set of data points (the \n\ IceCube data) and a population of candidate sources. 
The statistical variable of interest is the angular distance between each neutrino event and the candidate source closest to it. 
If the data and the potential sources are causally related, we expect an abundance in low distances. 
This forms the basic premise of our analysis. 

First, we define a dimensionless distance which is weighted by statistical spatial errors. 
Consider 
two objects in the sky with equatorial coordinates $(\alpha_i, \delta_i)$ and 
$ (\alpha_j, \delta_j)$, with the first coordinate being the Right Ascension (RA) and the second the declination (dec), and angular positional errors $\sigma_i$ and $\sigma_j$.\footnote{We assume that positional errors are symmetric in all directions, resulting in a spherical cone around each measured position. This is the case for the \n\ events considered here.} Their angular separation is then:
\begin{equation}
S_{ij} = \cos^{-1} \left( \sin{( \delta_i )} \sin{( \delta_j )} +  \cos{( \delta_i)} \cos{( \delta_j )} \cos{( \Delta \alpha_{ij} )} \right) ,
\end{equation}
where $\Delta \alpha_{ij}$ is the difference in right ascension coordinates, and their weighted, dimensionless angular distance can be defined as:
\begin{equation}
R_{ij} = \frac{ S_{ij} }{ \sqrt{\sigma_i^2 + \sigma_j^2} }~.  
\label{Rgeneraldef}
\end{equation}
The next step is to consider the set of $i=1,2, .....,N$ ($N=35$) neutrino data and $j=1,2,....,M$ candidates of a certain class, and, for each datum $i$, find the distance to the closest candidate\footnote{Here we are using the ``nearest neighbor" version of the method, which leads to identifying the nearest candidate as the most likely true source of a given \n.  This may lead to false attributions if more than one candidate overlaps with the \n\ data point within the error.  Considering the sparseness of the data and of the candidate sets we use, we estimate that the chance of false attribution be low. However, for future studies with larger samples, one may generalize the current method to include the distance to the second closest candidate as an additional statistical variable. }:
\begin{equation}
r_{i} = {\rm Min}_{\{j\}} R_{ij}~.  
\label{rdef}
\end{equation}
We then have $N$ values of $r$ -- the index $i$ will be dropped from here on out for simplicity of notation. 
In all cases considered in this work, the angular errors of the IceCube data dominate over the positional uncertainty of the sources, which therefore are neglected i.e., $\sigma_j=0$.  If a \n\ event has $r<1$, its  positional error encompasses 
the nearest candidate's on-sky location. In what follows, this condition will be used as an indicator of a plausible positional correlation between the datum and the candidate. Of particular interest will be the number of data for which the weighted distance is $r\leq1$.

The final steps consist of generating the distribution of the variable $r$ and comparing it with a null distribution. The latter is obtained through the hypothesis of a uniform distribution of sources in the sky. It represents the outcome expected if the data and the candidate sources are not causally related, and spatial coincidences are simply the result of random accident. The null distribution can be calculated, using a constant probability density for the sources. 
After some algebra (see appendix \ref{appA}), one gets: 
\begin{equation}
\frac{d{\mathcal P}(r)}{dr} =  \sum^{N}_{i=1} \sigma_i\frac{M}{2^M} \sin (r\sigma_i) \left[ 1+ \cos (r\sigma_i)\right]^{M-1}~.
\label{nulldist}
\end{equation}
Another approach -- which can be generalized to non-uniform populations of sources -- includes Monte Carlo simulations, in which we randomize the coordinates of the candidate sources
in both RA and dec.
Here the Monte Carlo method is used with $10^5$ iterations, averaging the distribution of $r$ 
over the number of iterations, so the resulting histogram is practically free from statistical errors associated with the finite number of candidate sources, $M$.  

When comparing the $r$-distribution of the data with the null one, the question to be answered is how compatible they are, i.e., how likely it is that the former might be a particular realization of the latter. 
To answer this question quantitatively, we use the $p$-value, defined as the probability that the null case produces, in one trial (see sec. \ref{sec:conclude} for a discussion of probabilities for multiple trials),  a number of coincidences ($r\leq 1$) equal to or larger than the one observed in the actual data.    Clearly, the larger the excess of the data over the null distribution at $r\leq1$, the lower the $p$-value.  Here the $p$-value is obtained by examining the $10^5$ Monte-Carlo-generated candidate source sets, and finding the percentage of these that have a number of coincidences equal to or exceeding the observed one at $r\leq1$. 

Finally, let us comment on the validity of this method and its underlying assumptions: \\
(i) no assumptions are made, nor needed, on the spatial distribution of the \n\ data set, and on if/how the data correlate spatially with one another.  Indeed (see appendix \ref{appA}), the main ingredient here is the probability to find a candidate source within the angular error of a given \n\ datum. We  have verified that our approach is valid for both uniform and non-uniform spatial distributions of the data in the sky.   \\
(ii) We stress that visually comparing the entire $r$-distribution with the randomized one (as shown in figures~\ref{fig:bz},~\ref{fig:sbg},~\ref{fig:sf}) only has indicative value.  This is because the histogram of the data is affected by large statistical errors associated with the small number of events in each bin and with the small number of candidates in each set. Therefore, we recommend relying mostly on the $p$-value as an indicator of compatibility.

%---------

\subsection{Combining neutrino events}
\label{sec:combine}
 
The larger the angular uncertainty of an IceCube neutrino event, the more difficult it becomes to disentangle its 
counterpart and provide a useful statistical evaluation of a population of candidates. We note that there are several instances of two or more \n\ shower events that are compatible, within the errors, 
with a common origin in space.  It might be of interest, then, to explore a common origin hypothesis, and treat the 
positionally overlapping events as different measurements of the position of the {\it assumed} same source. In this framework, 
the position of the source can be known more precisely by combining the multiple measurements into a single one, 
using the standard theory of measurement and errors.

We caution the reader that this hypothesis may imply un-physically
large neutrino fluxes for individual sources and therefore may be implausible. For this reason, the exercise of combining 
events should  be regarded as a useful check, that does not carry the same significance as an analysis involving all 
IceCube-reported neutrino
events. We only present the results of the ``combining" method  as a supplement to the more promising findings.

To combine the \n\ shower events, an iterative procedure is used.  At each iteration, all of the $N!$ possible pairs of data points are considered, with a weighted-distance, $C_{ij} = S_{ij} / (\sigma_i + \sigma_j)$.  If the lowest value of $C_{ij}$ is $C_{12}<1$ (i.e., the two data points have overlapping errors), the corresponding potential pair is combined into a single measurement of the position, with its resulting error, as follows:
\begin{eqnarray}
\mathrm{\alpha_c} = \frac{ \sum_{i=1}^{2} \alpha_i \cdot \sigma_i^{-2} }{ \sum_{i=1}^{2}\sigma_i^{-2} } \hskip 1truecm
\mathrm{\delta_c} = \frac{ \sum_{i=1}^{2}\delta_i \cdot \sigma_i^{-2} }{ \sum_{i=1}^{2} \sigma_i^{-2} } \hskip 1truecm \sigma_c^{-2} =  \sum_{i=1}^{2} \sigma_i^{-2}.
\end{eqnarray}
The new  neutrino position, ($\alpha_c, \delta_c$), and error, $\sigma_c$ -- which is smaller than either of the two initial errors, $\sigma_1$ and $\sigma_2$ -- are recorded and replace the original pair of events. 

The process is then repeated, until all overlapping neutrino events have been combined.
For an example of the ``combined" neutrino positions, see the all-sky plots in figures~\ref{fig:sbg} \&~\ref{fig:sf}.

%%%%%%%%%%%%%%%%%%%%%%%%%%%%%%%%

\section{Analysis of possible counterparts}
\label{sec:results}

In this section, the motivation for selected groups of sources and the results obtained are presented. 
Specifically, we give the distribution of the weighted distance, $r$,  (see section~\ref{sec:method}), the $p$-value, 
and the excess of events in the first bin ($r\leq 1$), $\Delta N$, relative to the null distribution.   
We provide, as an example, a test consistent with a null distribution to demonstrate a proof of method. However, we choose to focus mostly on the most interesting results found for groups of starburst galaxies and superbubble regions. A summary of the results for which a possible correlation is indicated is given in table~\ref{t:stats}. 

\subsection{``Null" results}
\label{sec:null}

\begin{figure}[tbp]
\centering
\includegraphics[width=0.6\textwidth]{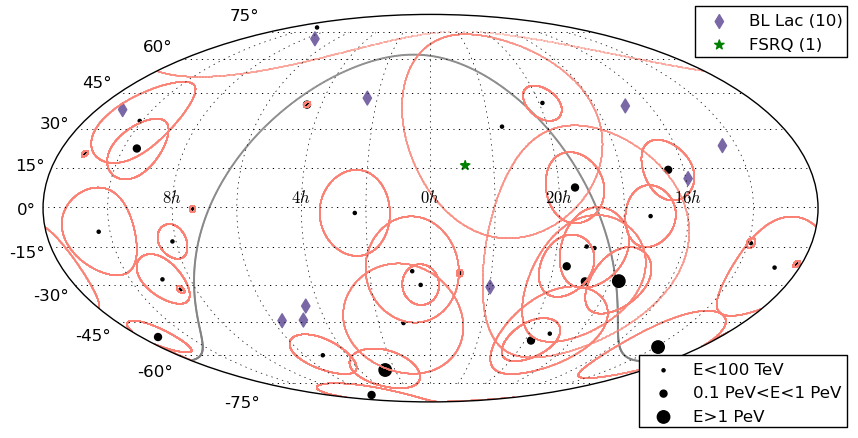}
\includegraphics[width=0.39\textwidth]{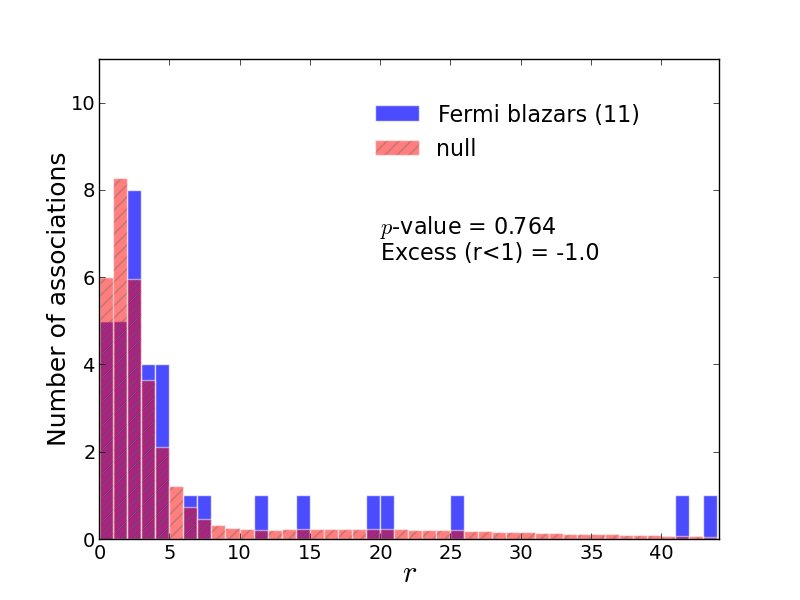}
\caption{Results for the eleven blazars from the $Fermi$-LAT 3FGL catalog, for which the 10--100 GeV  
gamma-ray flux density is $L_\gamma \geq 1 \times 10^{-9}$ photons cm$^{-2}$ s$^{-1}$.
\textit{Left:} The equatorial (J2000) coordinate sky map of these candidate sources; 
we distinguish between BL Lac objects  (light purple diamonds), and flat spectrum radio quasars (FSRQ; green stars).
The map also shows the 35 IceCube neutrino events as black dots, with their median angular error (pink ellipses). 
The dot size indicates the energy of the neutrino event (see legend). 
The solid grey line represents the galactic plane. 
\textit{Right:} The distribution of the weighted-distance to the nearest candidate source of the 35 neutrino events (solid, blue). 
The null distribution, determined via 10$^5$ iterations of a randomized population,
is also shown (pink with hash marks); purple indicates the overlap of the two histograms.
The legend gives the excess of the true distribution relative to the null in the the first bin ($r\geq 1$), and the $p$-value.}
\label{fig:bz}
\end{figure}

To underline the validity of our method, a test involving the brightest gamma-ray emitting blazars is shown in 
figure~\ref{fig:bz}. A complete description of the selection criteria for the blazar candidates can be found in section~\ref{appB}. 
Figure~\ref{fig:bz} shows, on the left, the location of the eleven
brightest sources on the sky and on the right, the resulting $r$ distribution of the neutrino events.
Overall, this distribution is consistent with the null case.  Five \n\ data points include a blazar within their median error, 
while six are expected in the null distribution.  The value $p\simeq $0.76  is found for the $p$-value (see also table~\ref{t:nullstats}), 
leading to the conclusion that there is no indication of a causal correlation between the
\ic\ neutrino events and the brightest blazars. This confirms the conclusions found with previous, 
positionally-blind selections of AGN \cite{Glusenkamp2015, Brown2015}.

A description of the analysis involving additional groups tested for which a ``null" result was obtained can be found appendix~\ref{appB}. Furthermore, a complete summary of the ``null" results can be found in table~\ref{t:nullstats} of appendix~\ref{appB}.

\subsection{Star-forming activity}
\label{sec:r-sf}

Galaxies undergoing star formation at high rates, $R_{SF}\simeq 10^{-1} - 10^2$ M\sun~yr$^{-1}$ \cite{Kennicutt1998}, 
usually caused by the disruption or a merger of galaxies, are known as star-forming galaxies. 
Galaxies with star formation rates up to $R_{SF} \simeq 20$ M\sun~yr$^{-1}$, or with typical supernova rates 
$R_{SN}\simeq$ 0.3 yr$^{-1}$, and more commonly observed
close-by hosting spiral disks, are referred to as starbursts. 
The emission observed over the entire electromagnetic spectrum of star-forming galaxies is dominated by
the evolutionary processes and environments of stars. A defining feature is their  luminous infrared emission, peaking
just short-ward of 100 $\mu$m, which is a product of dust absorbing UV radiation from massive stars and re-emitting it in the IR. 
Star-forming galaxies host large populations of objects associated with hadronic gamma-ray emission ---
including supernova remnants (SNR) \cite{Villante2008}, pulsar wind nebulae (PWN) \cite{Bednarek2003}, 
various types of explosions associated with supernova (SN) \cite{Senno2015, 	
Chakraborty2015, Asano2014, Razzaque2004, Razzaque2005}, and superbubbles  \cite{Bykov2014}
--- making them prime candidates for CR acceleration and high-energy neutrino emission
\citep{Loeb:2006tw}.
Much of this activity is connected to the interplay between massive stars and their surrounding media.

Massive stars ($M \gtrsim 8$ M\sun) form in dense molecular clouds and live relatively short lives 
($\sim 10^6$ yrs) before exploding as supernovae. They are typically observed in unbound, groups of ${\mathcal O}(10 -100) $
O and B stars, or ``OB associations" \cite{Blaauw1964}. The superimposed effects of their stellar winds and SN explosions
create giant ($> 150$ pc) cavities of hot, tenuous plasma, known as superbubbles \citep{McCray1979}.
Indeed, about 85\% of core-collapse SN occur in superbubbles, and starburst galaxies
each contain hundreds of these regions \citep{Higdon2005}.

Superbubbles are believed to be the origin of ``galactic fountains" \cite{Shapiro1976, Norman1989}. 
In this scenario, a star-forming region clears out 
surrounding gas and dust via stellar winds and SN explosions. As this region
grows, it will preferentially expand into lower density environments and, therefore, in a direction perpendicular to
the galactic plane.
In time, the star-forming region bursts through the galaxy's disk, exposing the gas and CRs to the halo. 
The strong magnetic fields contribute to this ``fountain" effect and direct the collimation of CRs outward.
This naturally leads to an amplification in CR acceleration \cite{Bykov2001, Parizot2004}.
Within the superbubble, particle acceleration may be affected by several different processes: 
shock acceleration in the winds, shock acceleration during SN explosions, and 
second order Fermi processes in the turbulent magnetic field deriving from merging stellar winds and SN ejecta \cite{Blasi2013, Bykov2014}. 

Since the detection of high-energy neutrinos, starburst galaxies have been studied as promising possible counterparts of IceCube sources
\citep{Fang2014, Anchordoqui2014, Ahlers2014, Wang2014, 	
Chakraborty2015, Senno2015, Chang2015, Liu2014, He2013, Tamborra2014, 
Murase2013}. 
These discussions largely rely on the rapid redshift evolution of star-forming galaxies (peaking during $z \sim 1-3$ \citep{Madau:1996aw}), 
implying that the majority of a diffuse high-energy neutrino flux should originate at redshift $z \gtrsim 1$.     
Sources at such large distances are difficult to resolve individually as \n\ point-sources, however, due to their low flux and high space density.
A positional association with two \ic\ events has been suggested in a study that  
 identified starburst galaxies as the primary 
candidates to produce the Telescope Array's UHECR 
($ E > 60$ TeV) excess  \citep{Fang2014}. 
Another interesting connection was the similarity between
the gamma-ray flux measured by $Fermi$-LAT within 20\deg~of the Galactic center and the IceCube neutrino flux
\citep{Razzaque2013}. The proposed scenario suggested increased star-forming activity coupled with CR confinement
by strong magnetic fields as the most likely origin of the gamma-ray and neutrino fluxes. Also within the inner galaxy, the young, compact star cluster Westerlund 1 appears to agree well with hard spectra, gamma-ray emission models and with the relative location of a number of IceCube neutrino events \citep{Bykov2015}. \\

In the following subsections, we present analyses for groups of 
(i) starburst galaxies and (ii) starburst galaxies plus local star-forming regions and superbubbles. 
More details on the individual candidates are given in section \ref{sec:discuss}.

\subsubsection{Starburst galaxies} 

\begin{figure}[tbp]
\centering
\includegraphics[width=0.6\textwidth]{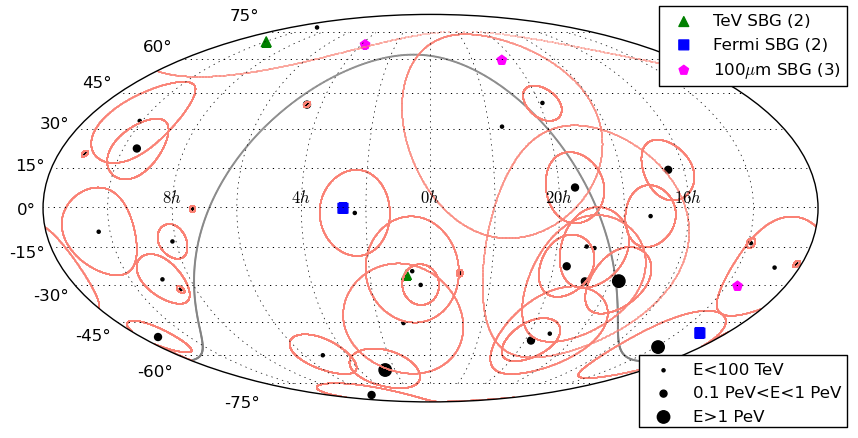}
\includegraphics[width=0.39\textwidth]{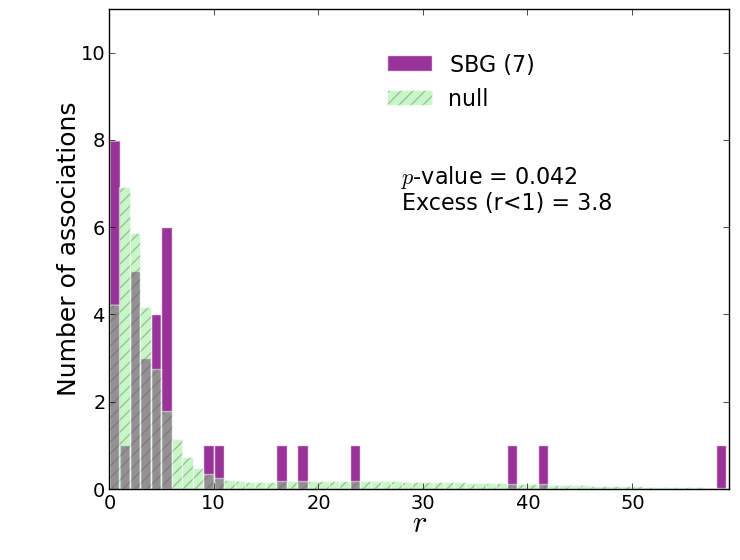}\\
\includegraphics[width=0.6\textwidth]{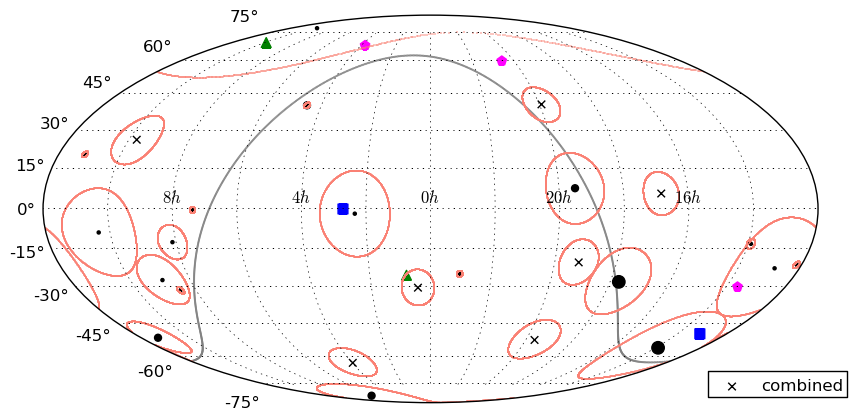}
\includegraphics[width=0.39\textwidth]{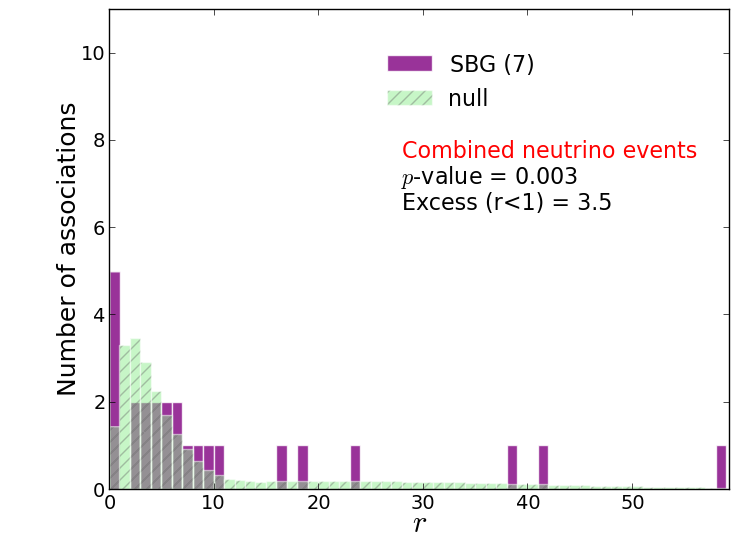}
\caption{
\textit{Top:}
The same as figure~\ref{fig:bz}, for all the starburst galaxies that pass the cut on the flux at 100 $\mu$m, S(100 $\mu$m) $\geq 250$ Jy \citep{Becker2009}.  We distinguish those that have been observed by $Fermi$-LAT (blue squares), and those that have been detected by both $Fermi$-LAT and in TeV gamma-rays (green triangles).  The latter, M 82 and NGC 253, are also the only starburst galaxies detected at TeV energies. 
The remaining starburst galaxies are shown as magenta pentagons. 
See table~\ref{t:obj} for the names and coordinates of each starburst.
\textit{Bottom:} Same as the top figures, but for the 25 ``combined" \n\ events dataset (see section~\ref{sec:combine}). The points that are the result of combining two or more events are shown as crosses. 
}
\label{fig:sbg}
\end{figure}

As a selection criterion for starburst galaxies, we made use of their emission at 100 $\mu$m, which is an indicator of star formation (see section \ref{sec:cat}). 
Specifically, a flux density cut S(100 $\mu$m) $\geq$ 250 Jy \cite{Becker2009, Sanders2003} was imposed, selecting seven 
starburst galaxies.  Two of these,  M 82 \cite{Acciari2009} and NGC 253
\cite{Acero:2009nb}, also have been detected in TeV gamma-ray energies and appear in the $Fermi$-LAT 3FGL catalog 
(classified as {\it sbg} there) as well. Two more of the seven objects appear in the $Fermi$-LAT 3FGL catalog, but have not been seen at TeV energies.

Figure~\ref{fig:sbg} shows the seven starburst galaxies on the sky map. All of them  lie within a neutrino
event error as reported by IceCube.  The $r$-distribution of the data has two peaks, one at $0\leq r\leq 1$ and another at $5\leq r\leq 6$.  This second peak is not expected in the  null distribution, and might be the result of a statistical fluctuation.  
The first peak has eight \n\ events, an excess $\Delta N = 3.8$ above the background or null distribution.  
This corresponds to a $p$-value of 0.042, indicating that there is only a 4\% probability that the outcome we find is obtained with uniformly distributed sources.  This is only moderately significant, but sufficiently interesting to motivate further investigation. 

In this spirit, in figure~\ref{fig:sbg} we also show the distribution of our ``combined" data set (see section
~\ref{sec:combine}). Five of these data overlap with a starburst galaxy within the positional error, constituting an excess of $\Delta N = 3.5$ relative to null, with a $p$-value of $0.003$.  Intuitively, the higher significance relative to the uncombined data set can be understood considering that the combined positional errors
cover a significantly smaller fraction of the sky, so that accidental (i.e., non-causal) coincidence is less likely.
The high significance should be contrasted, however, with potentially unphysical aspects of the combining exercise.  
Therefore, the most meaningful conclusions here are that, after combining overlapping neutrino data, an excess in the $r \leq 1$ persists, 
while the second peak at $r\sim 5$ does not, and therefore the former may be regarded as more robust.

In addition to the results shown in figure~\ref{fig:sbg}, in table~\ref{t:stats} we report further tests on the star formation hypothesis.
In one test, our method was applied to a reduced set of candidates, including only the four 3FGL starburst galaxies.
All of these lie within a neutrino event error, and the significance of the correlation is in-line with the result for the seven starbursts
($\Delta N=3.3$ and $p=0.046$ for the full data set, and $p=0.001$ for the combined one).

A second test was performed to include a deviation from the hypothesis of uniform candidate distribution. Indeed,
both gamma-ray and IR observations suffer from source confusion and a decreased sensitivity in the direction of the galactic plane \cite{Sanders2003, FermiLATCollaboration2014}. 
We attempted to account for this by excluding all the points that fall 
within $|b| \leq 10$\deg~of the galactic plane from the Monte Carlo-generated null distribution. 
The results of this test show a stronger correlation compared to figure~\ref{fig:sbg} for the IceCube neutrino
events ($p$-value of 0.034), as well as for the ``combined" events ($p$-value of $ 0.002$). In other words, allowing for the extragalactic Monte Carlo sources to avoid the galactic plane -- as the Fermi and IRAS detections do -- strengthens the conclusion of figure~\ref{fig:sbg}: extragalactic star-forming galaxies are a possible neutrino source.

Additional preliminary tests for which the randomization of the Monte Carlo was performed (i) by varying only the right ascension and not declination and (ii) with a distribution equivalent to the sensitivity of \ic, indicate that the randomizations done in both RA and declination produce the most conservative $p$-values. 

\begin{figure}[tbp]
\centering
\includegraphics[width=0.6\textwidth]{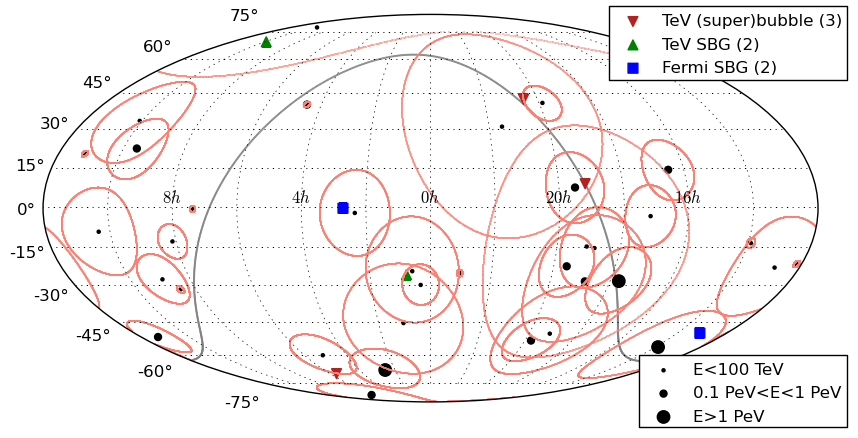}
\includegraphics[width=0.39\textwidth]{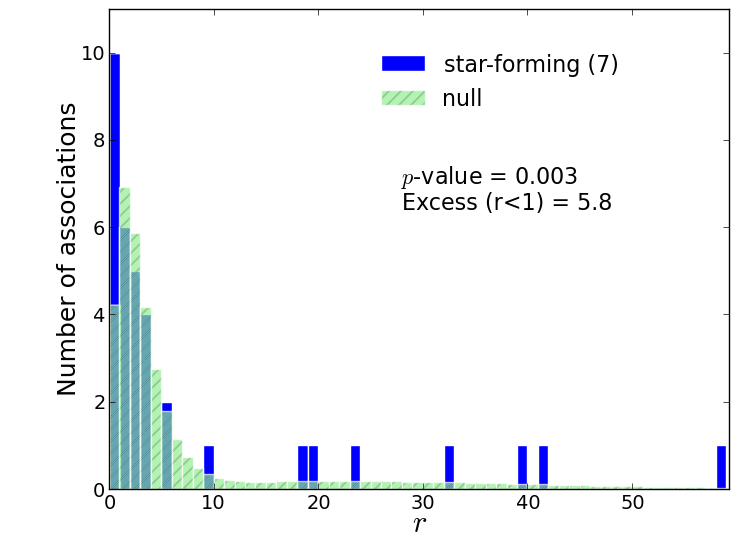}\\
\includegraphics[width=0.6\textwidth]{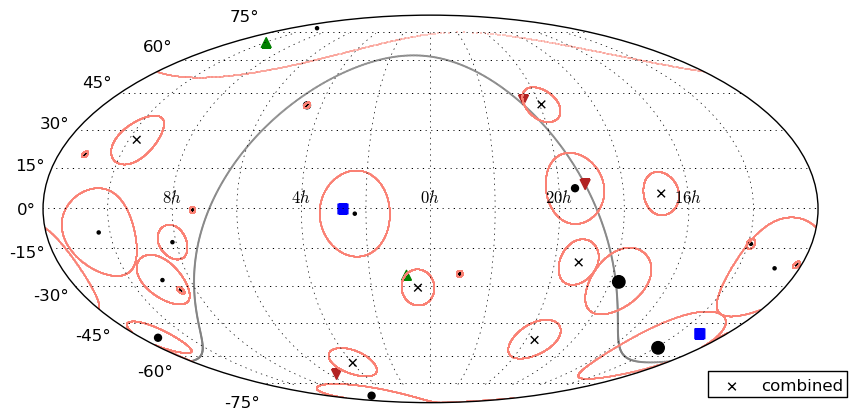}
\includegraphics[width=0.39\textwidth]{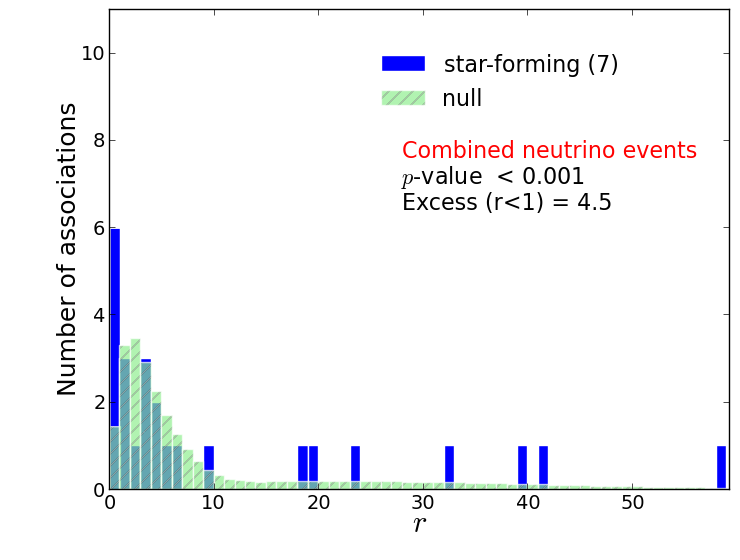}
\caption{The same as figure~\ref{fig:sbg}, for 
gamma-ray detected star-forming regions --- superbubbles and starburst galaxies.  In addition to 
TeV and $Fermi$-LAT detected
starburst galaxies (shown in figure~\ref{fig:sbg}), this analysis includes the superbubbles detected in TeV emission (down-pointing triangles).  
One of these objects, the Cygnus X Cocoon, was also reported in the 3FGL catalog. 
}
\label{fig:sf}
\end{figure}

\subsubsection{Starburst galaxies and local superbubble regions} 

Motivated by the interesting result for starburst galaxies of figure~\ref{fig:sbg}, let us now focus our attention more closely on local star-forming activity.  
We generated another set of candidate sources, with the criterion of supplementing the four gamma-ray detected starburst galaxies with superbubbles and star-forming regions that have been detected in TeV gamma rays. This addition amounts to two superbubbles and one bubble, or massive star-forming, region.\footnote{Note that two of the three added candidates are in our own Galaxy (see section \ref{sec:discuss}). Therefore, strictly speaking, the set of candidates sources in this subsection can not be a realization of the uniform distribution, which is a condition for the validity of the method used here. In this respect, the result obtained with only the four starburst galaxies (table \ref{t:stats}) can be considered more robust. }
One of the superbubbles, the Cygnus Cocoon, is also 
the only designated star-forming region in the 3FGL catalog. 

It was found (figure~\ref{fig:sf}) that ten \n\ events overlap with a candidate within their error ($r\leq 1$), amounting to an excess of
$\Delta N =5.8$ and a  $p$-value $ = 0.003$. This suggests a low degree of compatibility with the null case. A suggestion of a correlation improves further when considering the ``combined" neutrino events, where we find 6 positional associations with only $\sim1.5$ predicted, and the $p$-value falls below 0.001 (see table~\ref{t:stats} for a summary). \\

\begin{table}[tbp]
\centering
\begin{tabular}{|ccccccc|}
\hline
Candidate & Catalog(s)	&Selection	&Cand.		& count	& $\Delta N$  & $p$-value\\
		&			&Criteria 		&number		&($r\leq1$) &	    	&($r\leq1$)\\
\hline
\hline
		&			&			&			&		&		&  		\\
Starburst	&TeVCat, 3FGL &starburst	&4			&6		&3.3  &0.046 		\\
		&			&			&			&[4]		&[3.1]   &[0.001] 	\\
		&			&			&			&		&		&  		\\

\hline
		&			&			&			&		&		& 		\\
Starburst	&TeVCat,	3FGL & S$(100\mu m)\geq$ 250 Jy &7  	&8	&3.8 & 0.042 	 	\\
		&IRAS 100 $\mu$m &		&			&[5]	&[3.5] 	&[0.003] 	 	\\
		&			&			&			&		&		&  		\\

\hline
		&			&			&			&		&		&  		\\
Starburst	&TeVCat,	3FGL & same as above, &7		&8		&3.9   &0.034 		\\
		&IRAS 100 $\mu$m &randomize &			&[5]		&[3.6] 	&[0.002] 	\\
		&			&with $|b|>10$\deg &		&		&		&  		\\
		&			&			&			&		&		&  		\\

\hline
		&			&			&			&		&		&  		\\
Star-form.	&TeVCat, 3FGL &starburst,  	&7			&10	&5.8   	&0.003 		\\
		&			&superbubble, 	&			&[6]		&[4.5]  &$[<$0.001] 	\\
		&			&star-form. region &			&		&		&  		\\
		&			&			&			&		&		&  		\\
\hline
\end{tabular}
\caption{Summary of results presented in figures~\ref{fig:sbg} and ~\ref{fig:sf}, and additional tests
not shown in graphics. For each class of candidates, the table gives the number of objects in the sample (column 4). Entries in brackets refer to the combined data set, as explained in section~\ref{sec:combine}. Columns 5 and 6 give the number of neutrino associations with weighted distance $r \leq 1$ from the nearest  candidate and the excess over the number of associations expected in the null case, $\Delta N$, respectively. }
\label{t:stats}
\end{table}

In summary, in section~\ref{sec:results} candidate sources were examined related to AGN and/or star formation. While the former present a good compatibility with the null case, an interesting indication of deviation from it has emerged for star formation. While a robust claim is premature, this result is sufficiently interesting to prompt a number of tests to assess the plausibility of nearby star-forming sites as being the origin of a subset ($\sim 3- 6$ data points) of the \ic\ signal.  This is theme of the next section of this paper.

%%%%%%%%%%%%%%%%%%%%%%%%%%%%%%

\section{Discussion: star formation as a possible origin of high energy \ns}
\label{sec:discuss}

Let us take a closer look at the seven gamma-ray detected objects rich in star-forming activity that have emerged in section~\ref{sec:r-sf}, and apply naturalness considerations to estimate if they are plausible contributors to the \ic\ neutrino data.

\begin{figure}[tbp]
\centering
\includegraphics[width=0.7\textwidth]{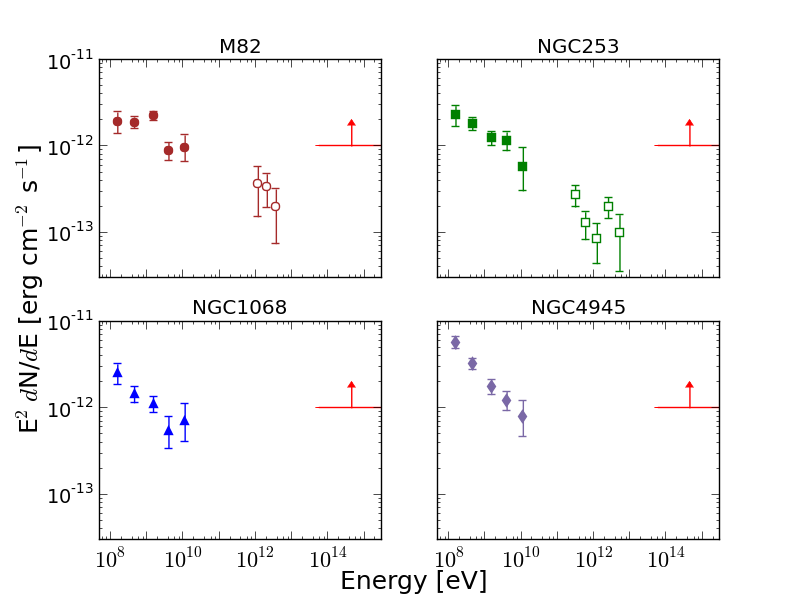} \\
\includegraphics[width=0.7\textwidth]{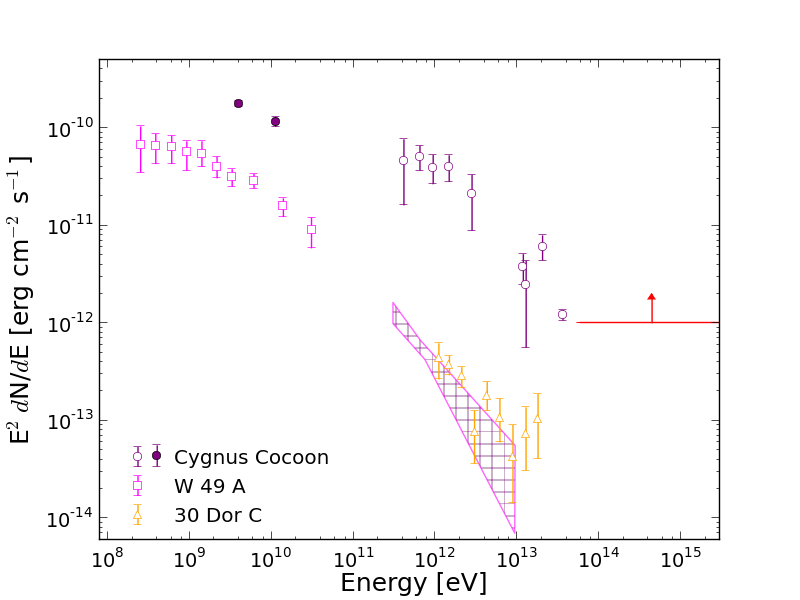}
\caption{Gamma-ray spectra of the candidate star-forming sources (figures~\ref{fig:sbg} and~\ref{fig:sf}, table \ref{t:obj})
that have been detected in gamma-rays.
The solid red line is an estimate of the minimum \n\ flux needed to produce one event at IceCube with significant likelihood (see text for details); the width of the line represents the energy window probed by IceCube.
{\it Top:} $Fermi$-LAT spectra of the four starburst classified galaxies in the 3FGL catalog \cite{FermiLATCollaboration2014}, 
shown as filled symbols. The
open symbols represent the VERITAS TeV data of M 82 \cite{Acciari2009} and the H.E.S.S. TeV data of NGC 253 \cite{Acero:2009nb}. 
{\it Bottom:} Same as top, for the nearby superbubbles and star-forming region. The Cygnus Cocoon TeV spectrum is from the ARGO-YBJ detector \citep{Bartoli:2014tqa}, while for W 49 A, the H.E.S.S and re-analyzed
$Fermi$-LAT GeV data are shown \cite{Brun2011}.  For 30 Dor C, the only data available are H.E.S.S TeV observations \cite{Abramowski:2015rca}.
}
\label{fig:spec}
\end{figure}

Table~\ref{t:obj} summarizes the main facts of these candidates --- and of those that have not been detected in gamma rays, which will not be discussed here --- and shows the ID number of the \n\ data that overlap with these to within the error. 
Their gamma ray spectra 
are shown in figure~\ref{fig:spec}. 
Below we describe each object individually. 

\begin{table}[tbp]
\centering
\begin{tabular}{|l|c|c|c|c|c|}
\hline
Name		&RA			&dec			&Class		&$D_L$	&$\nu$ ID \\ 
			&(J2000)		&(J2000)		&			& [Mpc] 	& \\ 			
\hline
\hline
NGC 253		&00 27 34		&$-$25 17 22	&starburst		& 3.1		& 7, 10, 21 \\ 
NGC 1068	&02 42 43		&$-$00 01 33	&starburst		& 13.7	& 1 \\ 
$[$IC 342$]$	&03 46 49		&+68 05 46	&starburst		& 4.6		& 31 \\ 
30 Dor C		&05 35 55		&$-$69 11 10	&superbubble	& 0.05	 & 19 \\ 	
M 82			&09 55 53		&+69 40 46	&starburst		& 3.6		& 31 \\ 
NGC 4945	&13 05 29		&$-$49 26 03	&starburst		& 3.9		& 35 \\ 
$[$M 83$]$	&13 37 01		&$-$29 51 57	&starburst		& 3.6		& 16 \\ 	
W 49 A		&19 10 27		&+09 11 25	&star-form region & 0.011 & 25, 33, 34 \\ 	
Cygnus Cocoon &20 28 41	&+41 10 12	&superbubble	& 0.002	& 29, 34 \\ 
$[$NGC 6946$] $ &20 34 52	&+60 09 13	&starburst		& 5.3		& 34 \\
\hline
\end{tabular}
\caption{The candidate star-forming sources considered in figures~\ref{fig:sbg} and~\ref{fig:sf}, with their equatorial coordinates
(columns 2 and 3). The names in brackets are those objects 
that have not been detected in gamma-rays, but appear among the brightest starbursts in the IRAS 100 $\mu$m catalog 
(see section \ref{sec:cat}). Column 5 gives the distances from Earth of each object, taken from \cite{Becker2009} for starbursts, \cite{Pietrzynski:2013gia} for 30 Dor C,  \cite{Gwinn1992} for W 49 A and  \cite{Hanson2003} for the Cygnus Cocoon. 
Also shown are the ID numbers (from \cite{Aartsen2014a}) of the \n\ events that have 
weighted distance $r\leq 1$ (eq. (\ref{rdef})) for each candidate.
}
\label{t:obj}
\end{table}

\paragraph{M 82} 
The nearly edge-on, starburst galaxy M 82 (NGC 3034), located $\sim$3.6 Mpc away, was the first starburst detected in TeV emission, 
and perhaps the first direct detection of an extragalactic source 
of hadronic gamma-ray emission \cite{Acciari2009, Abdo:2009aa}. It is the prototypical small starburst galaxy with 
an estimated supernova rate $R_{SN}\sim$0.1--0.3 yr$^{-1}$ \cite{Kronberg1985, Fenech2008}, a gamma-ray luminosity
$ L (> \mathrm{GeV}) \sim 2 \times 10^{40}$ erg s$^{-1}$ \cite{Lacki2011} and a photon number power-law index of $\Gamma = 2.21 \pm 0.06$ over the 100 MeV -- 100 GeV energy bands \cite{FermiLATCollaboration2014}. Interactions with 
neighboring galaxies, prominently the larger spiral M 81, have spurred star-forming activity, particularly in the central regions
\cite{Yun1994}.  

\paragraph{NGC 253} 
Thought of as a ``twin" of M 82, NGC 253 is similarly located at a distance of $\sim$3.1 Mpc 
with comparable infrared luminosity and
spectral distribution and is also seen approximately edge-on. 
The nucleus of NGC 253 contains a very active star-forming region 150 pc across \cite{Engelbracht1998}
in which SN occur at a rate  $R_{SN}\sim$ 0.1 yr$^{-1}$ \cite{Antonucci1988, Lenc2006}. GeV and TeV gamma-ray emission has been detected,
implying a luminosity of $L (> \mathrm{GeV}) \sim 5.6 \times 10^{39}$ erg s$^{-1}$ \cite{Lacki2011}, and a power-law fit 
with a photon index of $\Gamma = 2.34 \pm 0.003$ is consistent with no spectral break in the gamma-ray emission
\cite{Abramowski:2012xy}.
X-ray \cite{Fabbiano1984, Dahlem1998, Bauer2008}
and radio \cite{Carilli1992, Heesen2009} observations have revealed the presence of a hot diffuse halo resulting from
a ``disk wind" extending $\sim$9 kpc from the galactic plane. 

\paragraph{NGC 1068}
The most distant gamma-ray detected starburst galaxy, at $\sim 13.7$ Mpc,
NGC 1068 has the lowest detected 100 $\mu$m flux, yet its luminosity
is more than four times greater ($L_{100\,\mu\mathrm{m}} \sim 8.6 \times 10^{24}$ W Hz$^{-1}$) 
than the other starburst galaxies selected in our sample \cite{Ackermann:2012vca}. Indeed, it is the steep far-IR spectrum 
and its 100 $\mu$m luminosity 
that classifies this object as a starburst \cite{Sanders2003, FermiLATCollaboration2014}. 
However, its weak active nucleus, surrounded by a region of intense star formation extending $\sim$1 kpc \cite{Thronson1989}, 
has been widely discussed in the literature as the prototypical Seyfert. 
A study comparing the gamma-ray emission of NGC 1068 detected by $Fermi$-LAT with those of M 82, NGC 253, and 
NGC 4945 cited that its gamma-ray luminosity was too high to be explained only by starburst activity
\cite{Lenain2010}. The best-fit photon number power-law index for 100 MeV -- 100 GeV energies is $\Gamma = 2.32 \pm 0.10$
\cite{FermiLATCollaboration2014}.

\paragraph{NGC 4945} 
Also classified as both a Seyfert II and a starburst galaxy, NGC 4945 is a nearly edge-on barred spiral $\sim$3.9 Mpc distant.
Unlike NGC 1068, the high energy emission detected using $Fermi$-LAT could be explained only in terms
of its starburst activity \cite{Lenain2010}. NGC 4945 is one of the brightest 100 $\mu$m sources with a flux of 1330 Jy 
\cite{Sanders2003}, only slightly fainter than M 82. The best-fit photon number power-law index in the 100 MeV -- 100 GeV 
energy bands is $\Gamma = 2.43 \pm 0.07$ \cite{FermiLATCollaboration2014}.
A cone-shaped plume extending $\sim$500 pc from the nuclear region 
perpendicular to the disk has been detected in X-ray \cite{Schurch2002} and H$\alpha$ \cite{Rossa2003} and is believed to be 
driven by supernovae.

\paragraph{Cygnus Cocoon} 
The Cygnus Cocoon is a 50 pc wide star-forming region located in the Galaxy 1.5 kpc away \cite{Hanson2003}
with a combined mass of $\sim 8 \times 10^6$ M\sun~\citep{Ackermann2011}.
It hosts a collection of 1500-2000 massive OB stars \citep{LeDuigou2002} (e.g Cyg OB2), 
massive star clusters (e.g. NGC 6910), pulsars, 
SNRs (e.g. $\gamma$ Cygni), etc., whose combined effects from stellar winds and SN explosions have created 
a superbubble. Fifteen sources have been detected within the diffuse Cygnus Cocoon field in the 3FGL catalog, 
although the classification of most sources is currently unidentified and
some may be potentially confused with Galactic diffuse emission \citep{FermiLATCollaboration2014}.
The detection of the Cygnus Cocoon, using the $Fermi$-LAT (1--100 GeV) with a flux 
$L_\gamma \sim 6 \times 10^{-8}$ cm$^{-2}$ s$^{-1}$
and the Milagro Gamma-ray Observatory with a flux $L_\gamma \sim3.5 \times 10^{-11}$ cm$^{-2}$ s$^{-1}$ centered at 
$\sim$12 TeV \citep{Abdo:2006fq, Abdo:2007ad}, revealed a hard spectrum, most likely of diffuse, 
interstellar origin \citep{Ackermann2011}. 
The similarity between the emission morphology and IR/optical observed interstellar 
structure favors a CR origin, concluding that the Cygnus Cocoon is most likely a CR superbubble capable of accelerating CRs up
to an estimated 150 TeV \citep{Ackermann2011}. Additional detections at TeV energies have been confirmed by the 
 ARGO-YBJ detector \citep{Bartoli:2014tqa}.
 
\paragraph{W 49 A}
Also within the Galaxy, at a distance of 11.4 kpc \cite{Gwinn1992}, 
is the W 49 complex, hosting one of the most active and luminous ($L_{IR} > 10^7$ L\sun) star-forming regions in the Galaxy, 
W 49 A \cite{Sievers1991}. W 49 A contains $\sim$30 ultracompact HII regions \cite{Dreher1984, DePree1997}
for a total mass of $\sim10^6$ M\sun~\cite{Sievers1991}. Evidence for multiple expanding shells provided 
from the radiation pressure and/or strong stellar winds of massive stars has been observed ($\sim$15 pc scale), as well as the 
remnants of gas ejections on larger scales ($\sim 35 \times 15$ pc$^2$) \cite{Peng2010}.
W 49 A was detected in TeV gamma-rays at a 4.4$\sigma$ significance level, using H.E.S.S and re-analyzed
$Fermi$-LAT GeV data \cite{Brun2011}.

\paragraph{30 Doradus C}
30 Dor C is a 100 pc wide superbubble in the Large Megallanic Cloud, roughly 50 kpc away \cite{Pietrzynski:2013gia,Abramowski:2015rca}. 
It is luminous in radio, optical \cite{Mathewson1985},
X-ray (synchrotron-emitting) \cite{Bamba2004, Smith2004}
and TeV gamma-rays \cite{Abramowski:2015rca} with a central temperature of $7.4 \times 10^6$ K. 
Although neither leptonic or hadronic originating TeV gamma-ray emission
can be ruled out, conditions in the superbubble provide evidence for magnetic-field amplification combined with turbulence
in the hadronic scenario, possibly providing conditions for CRp acceleration exceeding energies of 3 PeV 
\cite{Abramowski:2015rca}.
\\

Next we will discuss if the physics of these star-forming objects supports strong \n\ emission. 
An immediate test of the promise of an object as a \ic\ source is to check if it is sufficiently powerful to produce at least one \n\ in \ic. This can be done using the observed gamma-ray spectra (figure~\ref{fig:spec}) under the assumption that \n\ and gamma-ray emission escaping the source are similar. This is expected to be true (i) when pair production can be neglected, (ii) in most applications of the hadronic model where the interstellar medium of the source is transparent to gamma-rays (for $N_{H} \lesssim10^{26}$ cm$^{-2}$ \cite{Lacki2012} at GeV energies), and (iii) the \n\ flux and spectrum trace the gamma-ray spectrum to within a factor $\sim 2-3$. 

For a \n\ spectrum of the form $E^{-2}$ 
the \n\ flux required to produce one event in \ic, $\phi_{(1)}$, is such that $E^2\phi_{(1)} \sim 10^{-11}~{\rm ergs~ cm^{-2}s^{-1}}$ \cite{Padovani2014}, nearly independently of the energy of the specific \n\ event considered, and with a Poissonian error of a factor of a few. As a conservative test of the plausibility of a neutrino source candidate, we require that its gamma-ray spectrum, power-law-extrapolated to the \ic\ energy window ($E\gtrsim 30$ TeV), be within an order of magnitude of $\phi_{(1)}$, i.e.  $E^2\phi_\gamma > E^2\phi_{min} \sim 10^{-12}~{\rm ergs~ cm^{-2}s^{-1}}$.  This criterion is similar to that in ref. \cite{Padovani2014}.

It appears that 
M 82 and NGC 253 (top two panels of figure~\ref{fig:spec}) do not pass this test, since their spectra decline rapidly with energy and their extrapolations fall below $\phi_{min} $ by at least one order of magnitude. A similar situation is realized for the remaining two starburst galaxies, NGC 1068 and NGC 4945. However, here the possibility of a hardening of the spectrum above a TeV remains open, due to the absence of observations in this regime. 
Slightly more promising are the spectra for the superbubbles/star-forming regions (bottom of figure~\ref{fig:spec}): while 30 Dor C and W 49 A also appear to fail the test, the Cygnus Cocoon passes.  
These results agree overall with past studies of starburst galaxies as high energy \n\ sources \cite{Romero:2003tj,Lacki:2010ue,Lacki2011,Yoast-Hull2014}, where, although with a strong dependence on the parameters, relatively soft \n\ spectra were found, reaching up to 150 TeV \cite{Lacki:2010ue}. 
Among the galactic objects, the Cygnus Cocoon has been predicted to be a detectable neutrino emitter 
\cite{Beacom2007, Gonzalez-Garcia2009, Fox2013}.

We found that the basic energetics test disfavors the simplest scenario of starburst galaxies with similar \n\ and gamma-ray spectra. However, it leaves open the possibility that the gamma-ray flux from starbursts is suppressed compared to the \n\ flux, due to unaccounted for absorption affecting the gamma-rays or their parent particles (e.g. ref. \cite{Murase2015}). The most likely possibility is the case of pair production, where the interaction between gamma-rays and a radiation field of lower energy photons produces electron-positron pairs. This mechanism is responsible for gamma-ray attenuation in connection with the extragalactic background light, and it effectively steepens gamma-ray spectra through absorption and also by redistributing gamma-ray photons to lower energies. 
The cross section for this interaction is maximized when the product of the photon energies
is $\sim0.1$ MeV$^{2}$. 
For example, for gamma-rays of 100 GeV the interaction is maximized with $\lambda \simeq 0.1$ $\mu$m UV photons and at 10 TeV energies this interaction is maximized for $\sim 10 $ $\mu$m mid-IR photons, both of which are abundant in these galaxies. 
Recent simulations have shown that gamma-rays above
$E \sim 2-5$ TeV have opacities above $\tau > 1$ in the star-forming galaxy ARP 220, M 82, and a similar but smaller effect in NGC 253 \cite{Torres2004, Lacki:2010ue, Yoast-Hull2015}. 

Without relying on gamma ray spectra, it is possible to test for compatibility between the subset of $n \sim 3-4$ events that might be due to the objects listed in table \ref{t:obj} --- which are all local, ($d < 15$ Mpc) --- and the whole set of  $N=35$ \n\ data at \ic\ of which $B\sim 13$ should be due to background (see section \ref{sec:data}).  
The argument is that the local contribution to the \n\ flux should
have a diffuse counterpart, due to similar objects at larger distances that can not be resolved individually.  
The ratio of local and diffuse fluxes can be calculated under the assumption that the local sources are  representative of most objects in their class, and using basic information on the cosmological evolution of the source population.  
We have estimated this ratio for a non-evolving population of identical objects, as well as one evolving like the cosmic star formation rate (see e.g., \cite{Hopkins:2006bw}), and we find that up to  $\sim 2\%$ of the total \n\ flux can be from objects with $d<15$ Mpc. In comparison, from the data we obtain a ratio $f = n/(N -B) \sim 0.14 - 0.18$.  

The difference between the two results can be interpreted as disfavoring the hypothesis of nearby starburst galaxies as \ic\ sources.  However, it could also indicate a positive fluctuation of the very nearby star formation rate with respect to the cosmological average, which has been suggested in connection with measurements of the nearby supernova rate \cite{Ando2005, Kistler2011, Mattila2012, Botticella2012}. Alternatively, it could indicate perhaps a mechanism whereby neutrinos 
are emitted preferentially along the plane of disk galaxies, 
leading to an enhanced contribution of edge-on galaxies to the total flux. We note in this context that 3 out of the 4 gamma-ray detected starburst galaxies in table~\ref{t:obj} are seen as edge-on, i.e., with an isophotal axis ratio of $b/a < 0.35$ \footnote{NASA/IPAC Extragalactic Database; http://ned.ipac.caltech.edu/ }, yet in complete samples of nearby and distant galaxies, on average only $\sim$15\% of the total galaxies are observed with this orientation \cite{Odewahn1997}. Finally, one should consider that exotic properties of the \n\ as a particle could 
increase the ratio $f$; for example, such an effect is predicted in models where a new \n\ interaction with a light mass mediator causes absorption of the cosmological contribution to the \n\ flux \cite{Cherry:2014xra}.

%%%%%%%%%%%%%%%%%%%%%%%%%%%%%%

\section{Summary and conclusion}
\label{sec:conclude}

We addressed the question of a possible contribution to the \ic\ data from local objects with powerful high energy emissions due to AGN and star-forming activities.  Specifically, we tested for a statistical positional correlation between the \n\ data and candidate sources from different catalogs, selecting among objects of the same class/morphology by
imposing a minimum flux in either gamma-rays or 100 $\mu$m emission.  The statistical quantity of interest is the weighted distance, $r$, between a \n\ data point and its nearest candidate source. We tested the distribution of $r$ against the ``null" hypothesis of random, non-causal, positional overlap with uniformly selected candidates sources.  

Results are consistent with a null hypothesis for control samples of nearby stars and exoplanets, as well as for blazars and Seyfert galaxies brightest in gamma-ray emission as seen in the $Fermi$-LAT 3FGL catalog (see figure~\ref{fig:bz} and appendix~\ref{appB}). 

More interesting results are found for objects with high star formation rates, such as starburst galaxies, superbubbles and massive star-forming regions.  The most significant excess at $r<1$ is found for the set of seven gamma ray-detected (from TeVCat and 3FGL catalogs) star-forming objects, including four starburst galaxies, two superbubbles (one galactic, the Cygnus Cocoon) and one galactic star-forming region.  Ten \n\ events overlap with a candidate within the error. The probability of this occurring in the null hypothesis for one trial ($p$-value) is $p=0.003$, as shown in figure~\ref{fig:sf}. 

 Similar excesses, although less significant, are seen for different selection criteria, which correspond to sets of candidate sources that partially overlap with one another.  In particular, the $p$-value is at the level of a few percent for the seven starburst galaxies that are brightest in far-IR 100 $\mu$m emission (see figure~\ref{fig:sbg}). This set includes the same four starburst galaxies as in the previous case.  If only the latter are considered, the $p$-value remains consistent with $p\simeq 0.05$. 
 
Taking a more global look at our results, one may wonder about the combined probability to obtain, in the null case, all the results shown here.  Statistically, the $K=8$ cases analyzed here (tables \ref{t:stats} and \ref{t:nullstats}) represent trials, each of them with a $p$-value, $p_i$ ($i=1,2,...,K$). Note that only $J=4$ trials (stars, blazars, Seyfert galaxies and starburst galaxies) are truly independent.  Let us focus on the minimum $p$-value obtained, $p_{min}=0.003$. 
A commonly used quantity \cite{Gross2010} is the probability to obtain at least one outcome with $p$-value $p_{min}$, over the number of trials done.  Restricted to the number of independent trials, one would get \cite{Choudalakis2011} $P=1-(1-p_{min})^J \simeq J p_{min} = 0.012$. $P$ is frequently referred to as the ``post-trial" $p$-value.   For the case at hand, of $K$ non-independent trials, a correct estimation of $P$ is more complicated and is beyond the scope of our paper.  However, generalizing the reasoning above, we can be confident that $J p_{min} \lta  P \lta  K p_{min}$, i.e., $0.012 \lta  P \lta  0.024$.

At this point, considerations on the significance of the excess of positional associations are necessarily mixed. On one hand, the excess is robust, since it appears for different candidate selection criteria and different models of the null case (uniform distribution or uniform with subtracted Galactic plane, see table \ref{t:stats}). In contrast, however, this result is not fully supported by a basic energetic test of neutrino emission tracing gamma-ray emission and the assumed redshift evolution of the star-forming population.

For the seven gamma-ray detected candidates, the gamma-ray spectra were examined. Under the assumption that  \n\ spectra  trace the gamma-ray ones, only one object, the Cygnus Cocoon in our galaxy, was found to be sufficiently powerful to produce one event in \ic\ with substantial likelihood. 
The remaining candidates can be reconciled with the observed excess in the hypothesis that their gamma ray flux might be substantially dimmer than the \n\ flux, due to e.g., high absorption. 
It is also worth noting that the excess corresponds to a local contribution (from sources within $\sim$15 Mpc distance) to the total \n\ flux of about 10--20\%. This is higher than the expected few percent when compared to the cosmological evolution of star formation, and could be explained by a local fluctuation in star-formation rate relative to the cosmic average or possibly, a preferential orientation for neutrino production in disk galaxies.

Looking ahead, we expect that the situation will become clearer with better statistics at \ic.  An update with 54 data points is upcoming (see \cite{botner2015} for preliminary presentations at conferences), and subsequent updates are expected at a rate of roughly 12 new events per year.  Therefore it is likely that, within a year or two, the excess we observe might either become disfavored, or confirmed with higher significance. A positive result would  have the character of discovery, and would be very fertile of theoretical developments on the physics of starburst galaxies and other star-forming regions. Questions to be investigated will be 
how opaque these star-forming regions could be to gamma-rays, and if the observational viewing angle of star-forming galaxies can affect the measurable neutrino flux.

%%%%%%%%%%%%%%%%%%%%%%%%%%%%%%

\acknowledgments

We thank Christopher Groppi and Soebur Razzaque for useful discussions and feedback. 
C.L. and K. E. acknowledge the National Science Foundation grant number PHY-1205745. C.L. is grateful to the Institute for Nuclear Theory of the University of Washington, Seattle, where this work was completed. 
R.W. is supported by NASA JWST grants NNX14AN10G and NAG5-12460.  

%%%%%%%%%%%%%%%%%%%%%%%%%%%%%%

\appendix
\section{Formalism: the null hypothesis}
\label{appA}

Here we derive the distribution of the weighted distance, $r$, for the null hypothesis, eq. (\ref{nulldist}). 
Let us consider a population of $M$ candidate sources, uniformly distributed in the sky, so that the probability to find a candidate in a unit of solid angle is $dp/d\Omega=1/(4\pi)$.  Consider now a generic point in the sky, with $\theta$ the angular distance from it. The probability to find a candidate at angular distance between $\theta$ and $\theta+d\theta$ is, then: 
\begin{equation}
dp(\theta) =\frac{1}{2} \sin \theta d\theta~. 
\label{diffp}
\end{equation}
By integration, one gets the probability to find a candidate at distance larger than $\theta$:
\begin{equation}
q(\theta)= \frac{1}{2}  (1+ \cos \theta)~.
\label{problarger}
\end{equation}
From these, we can obtain the probability that the nearest source is at angular distance between $\theta$ and $\theta+d\theta$. This is given by the probability that one source is between $\theta$ and $\theta+d\theta$, and all the other $M-1$ candidates are at a larger distance \cite{Sutherland1992}:
\begin{equation}
dP(\theta) = \frac{M}{2^M} \sin \theta (1+ \cos \theta)^{M-1} d\theta ~,
\label{ptot}
\end{equation}
where the factor $M$ in the numerator is found by the assumption of identical sources. 

Some observations on the distribution $dP/d\theta$ in eq. (\ref{ptot}):
\begin{itemize}

\item $dP/d\theta=0$ at $\theta=0$ and at $\theta=\pi$, as expected.  It has a maximum at $\theta=\theta_{max}$, with
\begin{equation}
\cos\theta_{max}=1-\frac{1}{M}~,
\label{polarmax}
\end{equation}
which agrees with the intuition that, for larger M, the most likely distance to the nearest source is smaller. 
 In the approximation  
$\theta \ll 1$, eq. (\ref{polarmax}) gives $\theta_{max} \simeq \sqrt{2/M}$. 
The dependence on $M^{-1/2}$ is expected, considering that for $M$ objects populating a two-dimensional space the area occupied by each object scales like $1/M$.  

\item The average of the distribution $dP/d\theta$ is found to be, in the limit $M\gg 1$:
\begin{equation}
\langle \theta \rangle \simeq \sqrt{\frac{\pi}{M}}~.
\label{polaravg}
\end{equation}

\end{itemize}

It is useful to express the distribution in eq. (\ref{ptot}) in terms of a weighted distance, $r=\theta/\sigma$, with $\sigma$ a constant (in our specific application, $\sigma$ is the angular error on the measured \n\ position, see section~\ref{sec:method}):
\begin{equation}
\frac{dP(r,\sigma)}{dr} = \sigma \frac{M}{2^M} \sin (r\sigma) \left[ 1+ \cos (r\sigma)\right]^{M-1}~.
\label{pnorm}
\end{equation}
For a set of $N$ \n\ data, each with error $\sigma_i$ ($i=1,2,.....,N$), the distribution of $r$ is the sum:
\begin{equation}
\frac{d{\mathcal P}(r)}{dr} = \sum^{N}_{i=1} \frac{dP(r,\sigma_i)}{dr} ~.
\label{rdist}
\end{equation}
The combination of eqs. (\ref{rdist}) and (\ref{pnorm}) gives the expression in eq. (\ref{nulldist}).  We have checked that this result coincides with the Monte Carlo-simulated one for a uniformly distributed population of candidates.

%%%%%%%%%%%%%%%%%%%%%%%%%%%%%%

\section{Null results}
\label{appB}

In this section, we present a statistical analysis of candidate sources tested that have produced results consistent with a null distribution. We choose two types of sources, the brightest stars and closest exoplanets, expected not to correlate with the location of \ic\ neutrino events, to show a confirmation of the validity of our method. We additionally present the results of the brightest emitting gamma-ray blazars and Seyfert galaxies. As done in section~\ref{sec:results}, we present the distribution of the weighted distance, r (see section~\ref{sec:method}), the $p$-value, and the excess of events in the first bin ($r \leq 1$), $\Delta N$, relative to the null distribution. A summary of null results is in table~\ref{t:nullstats}. 

%%%%%%%%%%

\paragraph{Stars}

We selected the seven brightest (in $m_V$) stars as seen in Smithsonian Astrophysical Observatory (SAO) Star Catalog \cite{SAO1966}, 
therefore having an apparent $V$ magnitude of $V \leq$ 0.3. Seven were chosen to ease computation and compare directly with the star-forming sources.
It was found that only two neutrino events overlapped with a stellar source resulting in a high likelihood, a p-value of 0.977, that
the positional coincidences of the brightest stars are consistent with the null distribution (see table~\ref{t:nullstats}).

%%%%%%%%%%

\paragraph{Exoplanets}

As an additional control sample, we found the seven closest stellar systems hosting exoplanets. To compile the ``candidate" sources, confirmed exoplanet systems from among the Exoplanet Orbit Database (EOD; \cite{Han2014}) were chosen to have a host star distance of $d \leq 4.95$ pc. Only one candidate source was considered for systems with more than one confirmed exoplanet. In this case, five neutrino events overlapped with a stellar system hosting an exoplanet, showing an excess of $\Delta N = 0.3$ over the expected 4.3 associations in the null distribution at $r \leq 1$, resulting in a $p = 0.530$ (see table~\ref{t:nullstats}).

\begin{table}[tbp]
\centering
\begin{tabular}{|ccccccc|}
\hline
Candidate & Catalog(s)	&Selection	&Cand.		& count	& $\Delta N$  & $p$-value\\
		&			&Criteria 		&number		&($r\leq1$) &	    	&($r\leq1$) \\
\hline
\hline
		&			&			&			&		&		&  \\
Blazar	&3FGL		&$F_{10-100\mathrm{GeV}} >$ &11 	&5		&--1.0&0.764 \\
		&			&$10^{-9}$ ph. cm$^{-2}$ s$^{-1}$ & 	&[1]		&[--1.2] &[0.938] \\
		&			&			&			&		&		&  \\
\hline
		&			&			&			&		&		&  \\
Seyfert	&3FGL		& Seyfert I \& II	&6			&6		&2.2  	&0.165 \\ 
		&			&			&			&[2]		&[0.7]  	&[0.368] \\
		&			&			&			&		&		& \\
				
\hline
		&			&			&			&		&		& \\
Stars		&SAO Star Cat	& $V \leq $ 0.3	& 7			& 2  		& -2.7	& 0.977 \\
		&			&			&			& [1]		& [-0.5] 	& [0.828] \\
		&			&			&			&		&		& \\

\hline
		&			&			&			&		&		&  \\
Exoplanets &EOD		& $d \leq 4.95$ pc & 7		& 5  		&  0.3 	& 0.530 \\
		&  			&			&			& [1]		& [-0.5]	& [0.828] \\
		&			&			&			&		&		& \\
\hline
\end{tabular}
\caption{The same as in table~\ref{t:stats} but for the ``null" results presented in figures~\ref{fig:bz},~\ref{fig:sey}, and from appendix~\ref{appB} not shown in graphics.}
\label{t:nullstats}
\end{table}

\paragraph{Blazars (AGN)}

Blazars are types of actively accreting AGN whose variable emission largely dominates their hosting galaxy, and 
for which highly relativistic beams are oriented along the line-of-sight \cite{Blandford1979, Urry1995}. 
They are divided into two major classes. Those displaying strong and broad optical emission lines are 
usually called flat-spectrum radio quasars (FSRQs), while objects with no broad emission lines are called 
BL Lacertae (BL Lac) objects
\cite{Giommi:2011sz}. Blazars are considered to be natural mechanisms for high-energy particle acceleration.
In particular, the acceleration of protons may explain:
(i) the energy transfer from the central engine over distances as large as 1 pc, (ii) the heating of a dusty disc in the nucleus
over distances of several 100 pc, and (iii) a near-infrared cut-off of the synchrotron emission in the jet \cite{Halzen1993}. 
However, recent literature \citep{Hinton2009, Holder2012} suggests that the currently favored mechanism for driving 
the high energy emission from blazars is a population of electrons accelerated to TeV energies, typically through 
Fermi acceleration by shocks in the AGN jet. TeV gamma-ray emission results from inverse Compton 
scattering off relativistic electrons, and the electrons cool by radiating X-ray synchrotron emission. The strong correlation
often observed between X-ray and TeV gamma-rays from blazars indicates a possibly common origin. 

In our analysis, we searched for the brightest AGN-classified objects in the 3FGL catalog. This included sources
such as blazars (both BL Lac \& FSRQ), AGN, radio galaxies, unidentified blazar candidates, and quasars
--- or those classified as {\it BLL, bll, FSRQ, fsrq, agn, RDG, rdg, BCU, bcu, ssrq} in 3FGL. 
A lower bound on the photon flux was imposed:   $L_\gamma > 10^{-9}$ photons cm$^{-2}$ s$^{-1}$ in the 10--100 GeV band.
Out of the 11 objects passing this criterium, 10 are BL Lac blazars and 1 is of FSRQ type. 
Indeed, a higher flux in gamma-ray emission is expected when the
orientation of the AGN jet is pointed in the direction of our viewing angle (blazars) compared with larger jet orientation
angles (radio galaxies, AGN classified, quasars, etc.) \cite{Hinton2009}. 
Figure~\ref{fig:bz} shows the resulting $r$ distribution of the neutrino events.
Overall, this distribution is consistent with the null case.  Five \n\ data points include a blazar within their median error, 
while six are expected in the null distribution.  The value $p\simeq $0.76  is found for the $p$-value, leading to the conclusion that there is no indication of a causal correlation between the
\ic\ neutrino events and the brightest blazars. This confirms the conclusions found with previous, 
positionally-blind selections of AGN \cite{Glusenkamp2015, Brown2015}. 

%%%%%%%%%%

\paragraph{Seyfert galaxies}

\begin{figure}[tbp]
\centering
\includegraphics[width=0.59\textwidth]{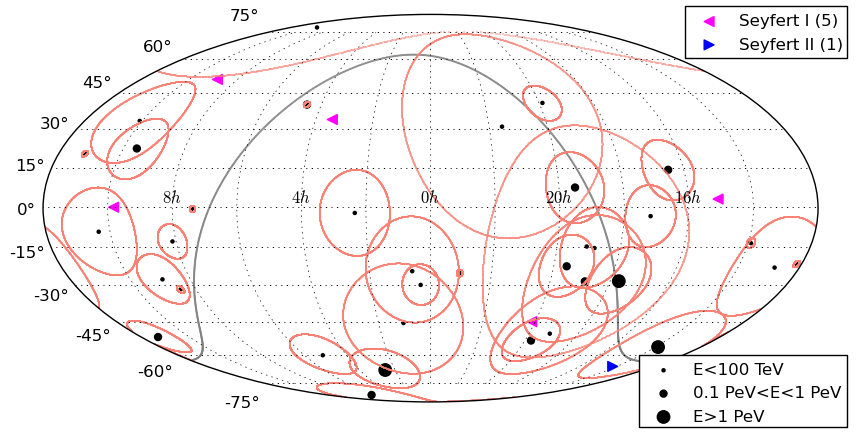}
\includegraphics[width=0.4\textwidth]{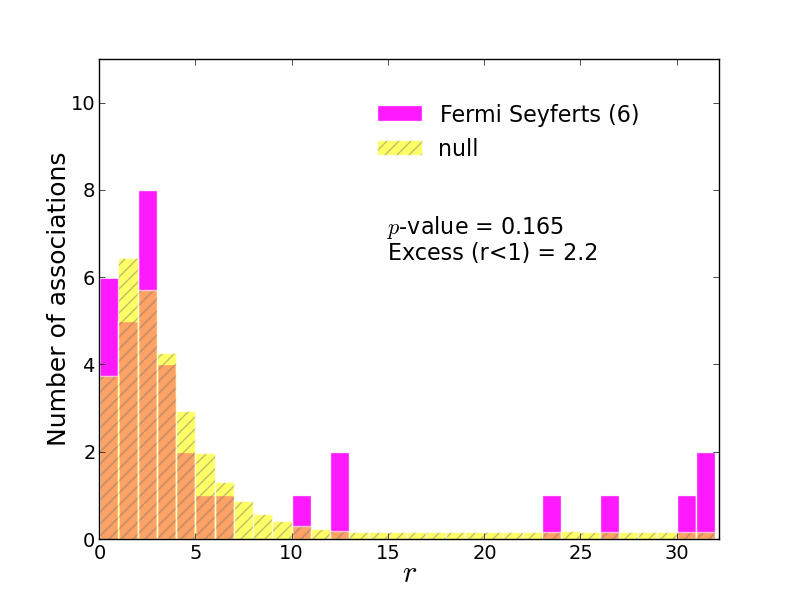}
\caption{The same graphics as figure~\ref{fig:bz}, now for the six Seyfert classified objects detected in the 3FGL catalog,
divided in Seyfert I objects (magenta left-pointing triangles) and Seyfert II objects (blue right-pointing triangles).}
\label{fig:sey}
\end{figure}

Seyfert classified galaxies are characterized by a bright nucleus, with an AGN strength in emission that is below that of a
quasar or blazar $L < 10^{11}$ L\sun\ \footnote{Here L\sun\ is the luminosity of the sun, 
L\sun\ $= 3.846 \times 10^{33}$ erg s$^{-1}$.} \cite{Seyfert1943, Schmidt1983}. 
Seyfert galaxies typically have spiral morphologies and active areas of star formation surrounding the nucleus.
Their emission has contributions from both the central AGN and star-forming activity in the galaxy disk. 
In the picture of AGN unification \citep{Urry1995}, Seyferts are identified as Type II when viewed edge on 
and Type I when the jet is oriented along the line-of-sight. 
Moreover, these AGN are relatively abundant at low $z$, as they are thought to be the evolutionary by-products of 
quieting quasars and blazars at higher redshifts \cite{Weedman1977}. 

For our analysis, we chose all six Seyfert classified objects in the $Fermi$-LAT 3FGL catalog.
Of these,  five are Seyfert I's (designated 
{\it NLSY1} or {\it nlsy1} in the 3FGL catalog) and one is Seyfert II (classified as {\it sey} in the 3FGL catalog).
Results from this analysis are shown in 
figure~\ref{fig:sey}. Three \ns\ events overlap with a candidate within the error. The excess in the first bin of the $r$-distribution 
is $\Delta N\simeq 2.2$, and a $p$-value of 0.165 is found. 
Hence, Seyferts do not constitute a significant signal; however, our results suggest that Seyfert galaxies may warrant further investigation.

%%%%%%%%%%%%%%%%%%%%%%%%%%%%%%

\bibliographystyle{JHEP}
\bibliography{icecube}

\providecommand{\href}[2]{#2}\begingroup\raggedright\begin{thebibliography}{100}

\bibitem{Aartsen2013b}
{IceCube Collaboration}, {\it {First Observation of PeV-Energy Neutrinos with
  IceCube}},  {\em Phys. Rev. Lett.} {\bf 111} (July, 2013) 021103,
  [\href{http://arxiv.org/abs/1304.5356}{{\tt arXiv:1304.5356}}].

\bibitem{Aartsen2013}
{IceCube Collaboration}, {\it {Evidence for high-energy extraterrestrial
  neutrinos at the IceCube detector.}},  {\em Science} {\bf 342} (Nov., 2013)
  1242856, [\href{http://arxiv.org/abs/1311.5238}{{\tt arXiv:1311.5238}}].

\bibitem{Aartsen2014a}
{IceCube Collaboration}, {\it {Observation of High-Energy Astrophysical
  Neutrinos in Three Years of IceCube Data}},  {\em Phys. Rev. Lett.} {\bf 113}
  (Sept., 2014) 101101, [\href{http://arxiv.org/abs/1405.5303}{{\tt
  arXiv:1405.5303}}].

\bibitem{Murase:2014tsa}
K.~Murase, {\it {On the Origin of High-Energy Cosmic Neutrinos}},
  \href{http://arxiv.org/abs/1410.3680}{{\tt arXiv:1410.3680}}.

\bibitem{Murase2013}
K.~Murase, M.~Ahlers, and B.~C. Lacki, {\it {Testing the hadronuclear origin of
  PeV neutrinos observed with IceCube}},  {\em Phys. Rev. D} {\bf 88} (Dec.,
  2013) 121301.

\bibitem{Liu2014}
R.-Y. Liu, X.-Y. Wang, S.~Inoue, R.~Crocker, and F.~Aharonian, {\it {Diffuse
  PeV neutrinos from EeV cosmic ray sources: Semirelativistic hypernova
  remnants in star-forming galaxies}},  {\em Phys. Rev. D} {\bf 89} (Apr.,
  2014) 083004.

\bibitem{Tamborra2014}
I.~Tamborra, S.~Ando, and K.~Murase, {\it {Star-forming galaxies as the origin
  of diffuse high-energy backgrounds: gamma-ray and neutrino connections, and
  implications for starburst history}},  {\em J. Cosmol. Astropart. Phys.} {\bf
  2014} (Sept., 2014) 043--043.

\bibitem{Chang2015}
X.-C. Chang, R.-Y. Liu, and X.-Y. Wang, {\it {Star-forming galaxies as the
  origin of the IceCube PeV neutrinos}},  {\em Astrophys. J.} {\bf 805} (May,
  2015) 95, [\href{http://arxiv.org/abs/1412.8361}{{\tt arXiv:1412.8361}}].

\bibitem{Loeb:2006tw}
A.~Loeb and E.~Waxman, {\it {The Cumulative background of high energy neutrinos
  from starburst galaxies}},  {\em JCAP} {\bf 0605} (2006) 003,
  [\href{http://arxiv.org/abs/astro-ph/0601695}{{\tt astro-ph/0601695}}].

\bibitem{Stecker:2006vz}
F.~W. Stecker, {\it {Are Diffuse High Energy Neutrinos from Starburst Galaxies
  Observable?}},  {\em Astropart.Phys.} {\bf 26} (2007) 398--401,
  [\href{http://arxiv.org/abs/astro-ph/0607197}{{\tt astro-ph/0607197}}].

\bibitem{Thompson:2006np}
T.~A. Thompson, E.~Quataert, E.~Waxman, and A.~Loeb, {\it {Assessing The
  Starburst Contribution to the Gamma-Ray and Neutrino Backgrounds}},
  \href{http://arxiv.org/abs/astro-ph/0608699}{{\tt astro-ph/0608699}}.

\bibitem{Aartsen2014b}
{IceCube Collaboration}, {\it {Searches for extended and point-like neutrino
  sources with four years of IceCube data}},  {\em Astrophys. J.} {\bf 796}
  (Nov., 2014) 109.

\bibitem{IceCubeCollaboration2014a}
{IceCube Collaboration}, {\it {Searches for small-scale anisotropies from
  neutrino point sources with three years of IceCube data}},
  \href{http://arxiv.org/abs/1408.0634}{{\tt arXiv:1408.0634}}.

\bibitem{Razzaque2013}
S.~Razzaque, {\it {Galactic Center origin of a subset of IceCube neutrino
  events}},  {\em Phys. Rev. D} {\bf 88} (Oct., 2013) 081302.

\bibitem{Bai2014}
Y.~Bai, A.~Barger, V.~Barger, R.~Lu, A.~Peterson, and J.~Salvado, {\it
  {Neutrino lighthouse at Sagittarius A*}},  {\em Phys. Rev. D} {\bf 90}
  (Sept., 2014) 063012.

\bibitem{Lunardini2014}
C.~Lunardini, S.~Razzaque, K.~T. Theodoseau, and L.~Yang, {\it {Neutrino events
  at IceCube and the Fermi bubbles}},  {\em Phys. Rev. D} {\bf 90} (July, 2014)
  023016.

\bibitem{Fang2014}
K.~Fang, T.~Fujii, T.~Linden, and A.~V. Olinto, {\it {Is the ultra-high energy
  cosmic-ray excess observed by the Telescope Array correlated with IceCube
  Neutrinos?}},  {\em Astrophys. J.} {\bf 794} (Oct., 2014) 126.

\bibitem{Moharana2015}
R.~Moharana and S.~Razzaque, {\it {Angular correlation of cosmic neutrinos with
  ultrahigh-energy cosmic rays and implications for their sources}},  {\em
  eprint arXiv:1501.05158} (2015).

\bibitem{Padovani2014}
P.~Padovani and E.~Resconi, {\it {Are both BL Lacs and pulsar wind nebulae the
  astrophysical counterparts of IceCube neutrino events?}},  {\em Mon. Not. R.
  Astron. Soc.} {\bf 443} (July, 2014) 474--484,
  [\href{http://arxiv.org/abs/1406.0376}{{\tt arXiv:1406.0376}}].

\bibitem{Sahu2014}
S.~Sahu and L.~S. Miranda, {\it {Some possible sources of IceCube TeV-PeV
  neutrino events}},  \href{http://arxiv.org/abs/1408.3664}{{\tt
  arXiv:1408.3664}}.

\bibitem{Krauss:2014tna}
F.~Krauss, M.~Kadler, K.~Mannheim, R.~Schulz, J.~TrŸstedt, et~al., {\it {TANAMI
  Blazars in the IceCube PeV Neutrino Fields}},  {\em Astron.Astrophys.} {\bf
  566} (2014) L7, [\href{http://arxiv.org/abs/1406.0645}{{\tt
  arXiv:1406.0645}}].

\bibitem{Krauss:2015pja}
{\bf Fermi-LAT} Collaboration, F.~Krauss et~al., {\it {TANAMI counterparts to
  IceCube high-energy neutrino events}},
  \href{http://arxiv.org/abs/1502.02147}{{\tt arXiv:1502.02147}}.

\bibitem{Petropoulou2015}
M.~Petropoulou, S.~Dimitrakoudis, P.~Padovani, A.~Mastichiadis, and E.~Resconi,
  {\it {Photohadronic origin of Formula-ray BL Lac emission: implications for
  IceCube neutrinos}},  {\em Mon. Not. R. Astron. Soc.} {\bf 448} (Mar., 2015)
  2412--2429, [\href{http://arxiv.org/abs/1501.0711}{{\tt arXiv:1501.0711}}].

\bibitem{ANTARES2015}
{ANTARES Collaboration}, {\it {ANTARES constrains a blazar origin of two
  IceCube PeV neutrino events}},  {\em Astron. Astrophys.} {\bf 576} (Mar.,
  2015) A8, [\href{http://arxiv.org/abs/1501.07843}{{\tt arXiv:1501.07843}}].

\bibitem{Brown2015}
A.~M. Brown, J.~Adams, and P.~M. Chadwick, {\it {Gamma-ray observations of
  extraterrestrial neutrino track event positions}},  {\em eprint
  arXiv:1505.00935} (2015).

\bibitem{Glusenkamp2015}
T.~Gl\"{u}senkamp and {for the IceCube Collaboration}, {\it {Analysis of the
  cumulative neutrino flux from Fermi-LAT blazar populations using 3 years of
  IceCube data}},  {\em eprint arXiv:1502.03104} (2015).

\bibitem{Anchordoqui2014}
L.~A. Anchordoqui, T.~C. Paul, L.~H.~M. da~Silva, D.~F. Torres, and B.~J.
  Vlcek, {\it {What IceCube data tell us about neutrino emission from
  star-forming galaxies (so far)}},  {\em Phys. Rev. D} {\bf 89} (June, 2014)
  127304, [\href{http://arxiv.org/abs/1405.7648}{{\tt arXiv:1405.7648}}].

\bibitem{Planck2015}
P.~Collaboration., {\it {Planck 2015 results. XIII. Cosmological parameters}},
  {\em eprint arXiv:1502.01589} (2015)
  [\href{http://arxiv.org/abs/1502.01589}{{\tt arXiv:1502.01589}}].

\bibitem{Hauser2001}
M.~G. Hauser and E.~Dwek, {\it {The Cosmic Infrared Background: Measurements
  and Implications 1}},  {\em Annu. Rev. Astron. Astrophys.} {\bf 39} (Sept.,
  2001) 249--307.

\bibitem{Atwood:2009ez}
{\bf Fermi-LAT} Collaboration, W.~Atwood et~al., {\it {The Large Area Telescope
  on the Fermi Gamma-ray Space Telescope Mission}},  {\em Astrophys.J.} {\bf
  697} (2009) 1071--1102, [\href{http://arxiv.org/abs/0902.1089}{{\tt
  arXiv:0902.1089}}].

\bibitem{FermiLATCollaboration2014}
{Fermi-LAT Collaboration}, {\it {Fermi Large Area Telescope Third Source
  Catalog}},  \href{http://arxiv.org/abs/1501.02003}{{\tt arXiv:1501.02003}}.

\bibitem{Becker2009}
J.~K. Becker, P.~L. Biermann, J.~Dreyer, and T.~M. Kneiske, {\it {Cosmic Rays
  VI - Starburst galaxies at multiwavelengths}},
  \href{http://arxiv.org/abs/0901.1775}{{\tt arXiv:0901.1775}}.

\bibitem{Sanders2003}
D.~B. Sanders, J.~M. Mazzarella, D.-C. Kim, J.~A. Surace, and B.~T. Soifer,
  {\it {The IRAS Revised Bright Galaxy Sample}},  {\em Astron. J.} {\bf 126}
  (Oct., 2003) 1607--1664.

\bibitem{Hinton2009}
J.~Hinton and W.~Hofmann, {\it {Teraelectronvolt Astronomy}},  {\em Annu. Rev.
  Astron. Astrophys.} {\bf 47} (Sept., 2009) 523--565.

\bibitem{Holder2012}
J.~Holder, {\it {TeV gamma-ray astronomy: A summary}},  {\em Astropart. Phys.}
  {\bf 39-40} (Dec., 2012) 61--75, [\href{http://arxiv.org/abs/1204.1267}{{\tt
  arXiv:1204.1267}}].

\bibitem{deRuiter1977}
H.~R. de~Ruiter, A.~G. Willis, and H.~C. Arp, {\it {A Westerbork 1415 MHz
  survey of background radio sources. II - Optical identifications with deep
  IIIA-J plates}},  {\em Astron. Astrophys. Suppl. Ser.} {\bf 28} (1977)
  211--293.

\bibitem{Windhorst1984}
R.~A. Windhorst, R.~G. Kron, and D.~C. Koo, {\it {A deep Westerbork survey of
  areas with multicolor Mayall 4 M plates. II - Optical identifications}},
  {\em Astron. Astrophys. Suppl. Ser. (ISSN 0365-0138)} {\bf 58} (1984) 39--87.

\bibitem{Sutherland1992}
W.~Sutherland and W.~Saunders, {\it {On the likelihood ratio for source
  identification}},  {\em Mon. Not. R. Astron. Soc. (ISSN 0035-8711)} {\bf 259}
  (1992) 413--420.

\bibitem{Virmani2002}
A.~Virmani, S.~Bhattacharya, P.~Jain, S.~Razzaque, J.~P. Ralston, and D.~W.
  Mckay, {\it {Angular correlation of ultra-high energy cosmic rays with
  compact radio-loud quasars}},  {\em Astropart. Phys.} {\bf 17} (July, 2002)
  489--495, [\href{http://arxiv.org/abs/astro-ph/0010235}{{\tt
  astro-ph/0010235}}].

\bibitem{Kennicutt1998}
R.~C. {Kennicutt, Jr.}, {\it {The Global Schmidt Law in Star‐forming
  Galaxies}},  {\em Astrophys. J.} {\bf 498} (May, 1998) 541--552.

\bibitem{Villante2008}
F.~L. Villante and F.~Vissani, {\it {How precisely can neutrino emission from
  supernova remnants be constrained by gamma ray observations?}},  {\em Phys.
  Rev. D} {\bf 78} (Nov., 2008) 103007.

\bibitem{Bednarek2003}
W.~Bednarek, {\it {Neutrinos from the pulsar wind nebulae}},  {\em Astron.
  Astrophys.} {\bf 407} (Aug., 2003) 1--6.

\bibitem{Senno2015}
N.~Senno, P.~M\'{e}sz\'{a}ros, K.~Murase, P.~Baerwald, and M.~J. Rees, {\it
  {Extragalactic star-forming galaxies with hypernovae and supernovae as
  high-energy neutrino and gamma-ray sources: the case of the 10 TeV neutrino
  data}},  {\em eprint arXiv:1501.04934} (Jan., 2015) 9,
  [\href{http://arxiv.org/abs/1501.04934}{{\tt arXiv:1501.04934}}].

\bibitem{Chakraborty2015}
S.~Chakraborty and I.~Izaguirre, {\it {Diffuse neutrinos from extragalactic
  supernova remnants: Dominating the 100 TeV IceCube flux}},  {\em Phys. Lett.
  B} {\bf 745} (2015) 35--39, [\href{http://arxiv.org/abs/1501.02615}{{\tt
  arXiv:1501.02615}}].

\bibitem{Asano2014}
K.~Asano and P.~M\'{e}sz\'{a}ros, {\it Neutrino and cosmic-ray release from
  gamma-ray bursts: Time-dependent simulations},  {\em Astrophys. J.} {\bf 785}
  (Apr., 2014) 54.

\bibitem{Razzaque2004}
S.~Razzaque, P.~M\'esz\'aros, and E.~Waxman, {\it {TeV Neutrinos from Core
  Collapse Supernovae and Hypernovae}},  {\em Phys. Rev. Lett.} {\bf 93}
  (2004), no.~18 181101, [\href{http://arxiv.org/abs/astro-ph/0407064}{{\tt
  astro-ph/0407064}}].

\bibitem{Razzaque2005}
S.~Razzaque, P.~M\'esz\'aros, and E.~Waxman, {\it {High Energy Neutrinos from a
  Slow Jet Model of Core Collapse Supernovae}},  {\em Modern Phys. Lett. A}
  {\bf 20} (2005), no.~31 2351--2367,
  [\href{http://arxiv.org/abs/astro-ph/0509729}{{\tt astro-ph/0509729}}].

\bibitem{Bykov2014}
A.~M. Bykov, {\it {Nonthermal particles and photons in starburst regions and
  superbubbles}},  {\em Astron. Astrophys. Rev.} {\bf 22} (2014) 77.

\bibitem{Blaauw1964}
A.~Blaauw, {\it {The O Associations in the Solar Neighborhood}},  {\em Annu.
  Rev. Astron. Astrophys.} {\bf 2} (Sept., 1964) 213--246.

\bibitem{McCray1979}
R.~McCray and T.~P. Snow, {\it {The Violent Interstellar Medium}},  {\em Annu.
  Rev. Astron. Astrophys.} {\bf 17} (Sept., 1979) 213--240.

\bibitem{Higdon2005}
J.~C. Higdon and R.~E. Lingenfelter, {\it {OB Associations,
  Supernova‐generated Superbubbles, and the Source of Cosmic Rays}},  {\em
  Astrophys. J.} {\bf 628} (Aug., 2005) 738--749.

\bibitem{Shapiro1976}
P.~R. Shapiro and G.~B. Field, {\it {Consequences of a New Hot Component of the
  Interstellar Medium}},  {\em Astrophys. J.} {\bf 205} (May, 1976) 762.

\bibitem{Norman1989}
C.~A. Norman and S.~Ikeuchi, {\it {The disk-halo interaction - Superbubbles and
  the structure of the interstellar medium}},  {\em Astrophys. J.} {\bf 345}
  (Oct., 1989) 372.

\bibitem{Bykov2001}
A.~M. Bykov and I.~N. Toptygin, {\it {A model of particle acceleration to high
  energies by multiple supernova explosions in OB associations}},  {\em Astron.
  Lett.} {\bf 27} (Oct., 2001) 625--633.

\bibitem{Parizot2004}
E.~Parizot, A.~Marcowith, E.~van~der Swaluw, A.~M. Bykov, and V.~Tatischeff,
  {\it {Superbubbles and energetic particles in the Galaxy}},  {\em Astron.
  Astrophys.} {\bf 424} (Sept., 2004) 747--760.

\bibitem{Blasi2013}
P.~Blasi, {\it {The origin of galactic cosmic rays}},  {\em Astron. Astrophys.
  Rev.} {\bf 21} (Nov., 2013) 70.

\bibitem{Ahlers2014}
M.~Ahlers and F.~Halzen, {\it {Pinpointing extragalactic neutrino sources in
  light of recent IceCube observations}},  {\em Phys. Rev. D} {\bf 90} (Aug.,
  2014) 043005.

\bibitem{Wang2014}
B.~Wang, X.~Zhao, and Z.~Li, {\it {Implications of Fermi -LAT observations on
  the origin of IceCube neutrinos}},  {\em J. Cosmol. Astropart. Phys.} {\bf
  2014} (Nov., 2014) 028--028, [\href{http://arxiv.org/abs/1407.2536}{{\tt
  arXiv:1407.2536}}].

\bibitem{He2013}
H.-N. He, T.~Wang, Y.-Z. Fan, S.-M. Liu, and D.-M. Wei, {\it {Diffuse PeV
  neutrino emission from ultraluminous infrared galaxies}},  {\em Phys. Rev. D}
  {\bf 87} (Mar., 2013) 063011.

\bibitem{Madau:1996aw}
P.~Madau, H.~C. Ferguson, M.~E. Dickinson, M.~Giavalisco, C.~C. Steidel,
  et~al., {\it {High redshift galaxies in the hubble deep field. color
  selection and star formation history to z=4}},
  \href{http://arxiv.org/abs/astro-ph/9607172}{{\tt astro-ph/9607172}}.

\bibitem{Bykov2015}
A.~M. Bykov, D.~Ellison, P.~Gladilin, and S.~Osipov, {\it {Ultra-hard spectra
  of PeV neutrinos from supernovae in compact star clusters}},  {\em eprint
  arXiv:1507.04018} (2015).

\bibitem{Acciari2009}
{VERITAS Collaboration}, {\it {A connection between star formation activity and
  cosmic rays in the starburst galaxy M82}},  {\em Nature} {\bf 462} (Nov.,
  2009) 770--772, [\href{http://arxiv.org/abs/0911.0873}{{\tt
  arXiv:0911.0873}}].

\bibitem{Acero:2009nb}
{\bf HESS} Collaboration, .~F. Acero, {\it {Detection of Gamma Rays From a
  Starburst Galaxy}},  {\em Science} {\bf 326} (2009) 1080,
  [\href{http://arxiv.org/abs/0909.4651}{{\tt arXiv:0909.4651}}].

\bibitem{Pietrzynski:2013gia}
G.~Pietrzynski, D.~Graczyk, W.~Gieren, I.~Thompson, B.~Pilecki, et~al., {\it
  {An eclipsing binary distance to the Large Magellanic Cloud accurate to 2 per
  cent}},  {\em Nature} {\bf 495} (2013) 76--79,
  [\href{http://arxiv.org/abs/1303.2063}{{\tt arXiv:1303.2063}}].

\bibitem{Gwinn1992}
C.~R. Gwinn, J.~M. Moran, and M.~J. Reid, {\it {Distance and kinematics of the
  W49N H2O maser outflow}},  {\em Astrophys. J.} {\bf 393} (July, 1992) 149.

\bibitem{Hanson2003}
M.~M. Hanson, {\it {A Study of Cygnus OB2: Pointing the Way toward Finding Our
  Galaxy's Super–Star Clusters}},  {\em Astrophys. J.} {\bf 597} (Nov.,
  2003) 957--969.

\bibitem{Abdo:2009aa}
{\bf Fermi-LAT} Collaboration, A.~Abdo, {\it {Detection of Gamma-Ray Emission
  from the Starburst Galaxies M82 and NGC 253 with the Large Area Telescope on
  Fermi}},  {\em Astrophys.J.} {\bf 709} (2010) L152--L157,
  [\href{http://arxiv.org/abs/0911.5327}{{\tt arXiv:0911.5327}}].

\bibitem{Kronberg1985}
P.~P. Kronberg, P.~Biermann, and F.~R. Schwab, {\it {The nucleus of M82 at
  radio and X-ray bands - Discovery of a new radio population of supernova
  candidates}},  {\em Astrophys. J.} {\bf 291} (Apr., 1985) 693.

\bibitem{Fenech2008}
D.~M. Fenech, T.~W.~B. Muxlow, R.~J. Beswick, A.~Pedlar, and M.~K. Argo, {\it
  {Deep MERLIN 5 GHz radio imaging of supernova remnants in the M82
  starburst}},  {\em Mon. Not. R. Astron. Soc.} {\bf 391} (Dec., 2008)
  1384--1402.

\bibitem{Lacki2011}
B.~C. Lacki, T.~A. Thompson, E.~Quataert, A.~Loeb, and E.~Waxman, {\it {On the
  GeV and TeV detections of the starburst galaxies M82 and NGC 253}},  {\em
  Astrophys. J.} {\bf 734} (June, 2011) 107.

\bibitem{Yun1994}
M.~S. Yun, P.~T. Ho, and K.~Y. Lo, {\it {A high-resolution image of atomic
  hydrogen in the M81 group of galaxies.}},  {\em Nature} {\bf 372} (Dec.,
  1994) 530--2.

\bibitem{Engelbracht1998}
C.~W. Engelbracht, M.~J. Rieke, G.~H. Rieke, D.~M. Kelly, and J.~M. Achtermann,
  {\it {The Nuclear Starburst in NGC 253}},  {\em Astrophys. J.} {\bf 505}
  (Oct., 1998) 639--658.

\bibitem{Antonucci1988}
R.~R.~J. Antonucci and J.~S. Ulvestad, {\it {A large family of compact radio
  sources in the starburst nucleus of NGC 253}},  {\em Astrophys. J.} {\bf 330}
  (July, 1988) L97.

\bibitem{Lenc2006}
E.~Lenc and S.~J. Tingay, {\it {The Subparsec-Scale Radio Properties of
  Southern Starburst Galaxies. I. Supernova Remnants, the Supernova Rate, and
  the Ionized Medium in the NGC 253 Starburst}},  {\em Astron. J.} {\bf 132}
  (Sept., 2006) 1333--1345.

\bibitem{Abramowski:2012xy}
{\bf HESS} Collaboration, A.~Abramowski et~al., {\it {Spectral analysis and
  interpretation of the $\gamma$-ray emission from the Starburst galaxy NGC
  253}},  {\em Astrophys.J.} {\bf 757} (2012) 158,
  [\href{http://arxiv.org/abs/1205.5485}{{\tt arXiv:1205.5485}}].

\bibitem{Fabbiano1984}
G.~Fabbiano and G.~Trinchieri, {\it {The complex X-ray emission of NGC 253}},
  {\em Astrophys. J.} {\bf 286} (Nov., 1984) 491.

\bibitem{Dahlem1998}
M.~Dahlem, K.~A. Weaver, and T.~M. Heckman, {\it {An X‐Ray Minisurvey of
  Nearby Edge‐on Starburst Galaxies. I. The Data}},  {\em Astrophys. J.
  Suppl. Ser.} {\bf 118} (Oct., 1998) 401--453.

\bibitem{Bauer2008}
M.~Bauer, W.~Pietsch, G.~Trinchieri, D.~Breitschwerdt, M.~Ehle, M.~J. Freyberg,
  and A.~M. Read, {\it {XMM-Newton observations of the diffuse X-ray emission
  in the starburst galaxy NGC~253}},  {\em Astron. Astrophys.} {\bf 489} (Oct.,
  2008) 1029--1046.

\bibitem{Carilli1992}
C.~L. Carilli, M.~A. Holdaway, P.~T.~P. Ho, and C.~G. de~Pree, {\it {Discovery
  of a synchrotron-emitting halo around NGC 253}},  {\em Astrophys. J.} {\bf
  399} (Nov., 1992) L59.

\bibitem{Heesen2009}
V.~Heesen, R.~Beck, M.~Krause, and R.-J. Dettmar, {\it {Cosmic rays and the
  magnetic field in the nearby starburst galaxy NGC 253}},  {\em Astron.
  Astrophys.} {\bf 494} (Feb., 2009) 563--577.

\bibitem{Ackermann:2012vca}
{\bf Fermi-LAT} Collaboration, M.~Ackermann et~al., {\it {GeV Observations of
  Star-forming Galaxies with \textit{Fermi} LAT}},  {\em Astrophys.J.} {\bf
  755} (2012) 164, [\href{http://arxiv.org/abs/1206.1346}{{\tt
  arXiv:1206.1346}}].

\bibitem{Thronson1989}
J.~Thronson, Harley~A., M.~Hereld, S.~Majewski, M.~Greenhouse, P.~Johnson,
  E.~Spillar, C.~E. Woodward, D.~A. Harper, and B.~J. Rauscher, {\it
  {Near-infrared image of NGC 1068 - Bar-driven star formation and the
  circumnuclear composition}},  {\em Astrophys. J.} {\bf 343} (Aug., 1989) 158.

\bibitem{Lenain2010}
J.-P. Lenain, C.~Ricci, M.~T\"{u}rler, D.~Dorner, and R.~Walter, {\it
  {Seyfert~2 galaxies in the GeV band: jets and starburst}},  {\em Astron.
  Astrophys.} {\bf 524} (Nov., 2010) A72,
  [\href{http://arxiv.org/abs/1008.5164}{{\tt arXiv:1008.5164}}].

\bibitem{Schurch2002}
N.~J. Schurch, T.~P. Roberts, and R.~S. Warwick, {\it {High-resolution X-ray
  imaging and spectroscopy of the core of NGC 4945 with XMM-Newton and
  Chandra}},  {\em Mon. Not. R. Astron. Soc.} {\bf 335} (Sept., 2002) 241--246.

\bibitem{Rossa2003}
J.~Rossa and R.-J. Dettmar, {\it {An H?�survey aiming at the detection of
  extraplanar diffuse ionized gas in halos of edge?on spiral galaxies}},  {\em
  Astron. Astrophys.} {\bf 406} (Aug., 2003) 505--525.

\bibitem{Ackermann2011}
M.~Ackermann et~al., {\it {A cocoon of freshly accelerated cosmic rays detected
  by Fermi in the Cygnus superbubble.}},  {\em Science} {\bf 334} (Nov., 2011)
  1103--7.

\bibitem{LeDuigou2002}
J.-M. {Le Duigou} and J.~Kn�dlseder, {\it {Characteristics of new star
  cluster candidates in the Cygnus area}},  {\em Astron. Astrophys.} {\bf 392}
  (Sept., 2002) 869--884.

\bibitem{Abdo:2006fq}
A.~Abdo, B.~T. Allen, D.~Berley, E.~Blaufuss, S.~Casanova, et~al., {\it
  {Discovery of TeV Gamma-Ray Emission from the Cygnus Region of the Galaxy}},
  {\em Astrophys.J.} {\bf 658} (2007) L33--L36,
  [\href{http://arxiv.org/abs/astro-ph/0611691}{{\tt astro-ph/0611691}}].

\bibitem{Abdo:2007ad}
A.~Abdo, B.~T. Allen, D.~Berley, S.~Casanova, C.~Chen, et~al., {\it {TeV
  Gamma-Ray Sources from a Survey of the Galactic Plane with Milagro}},  {\em
  Astrophys.J.} {\bf 664} (2007) L91--L94,
  [\href{http://arxiv.org/abs/0705.0707}{{\tt arXiv:0705.0707}}].

\bibitem{Bartoli:2014tqa}
{\bf ARGO-YBJ} Collaboration, B.~Bartoli et~al., {\it {Identification of the
  TeV Gamma-ray Source ARGO J2031+4157 with the Cygnus Cocoon}},  {\em
  Astrophys.J.} {\bf 790} (2014), no.~2 152,
  [\href{http://arxiv.org/abs/1406.6436}{{\tt arXiv:1406.6436}}].

\bibitem{Sievers1991}
A.~W. Sievers, P.~G. Mezger, M.~A. Bordeon, E.~Kreysa, C.~G.~T. Haslam, and
  R.~Lemke, {\it {Dust emission from star forming regions. I - The W49A and
  W51A complexes}},  {\em Astron. Astrophys. (ISSN 0004-6361)} {\bf 251} (1991)
  231--244.

\bibitem{Dreher1984}
J.~W. Dreher, K.~J. Johnston, W.~J. Welch, and R.~C. Walker, {\it {Ultracompact
  structure in the H II region W49N}},  {\em Astrophys. J.} {\bf 283} (Aug.,
  1984) 632.

\bibitem{DePree1997}
C.~G. {De~Pree}, D.~M. Mehringer, and W.~M. Goss, {\it {Multifrequency,
  High-Resolution Radio Recombination Line Observations of the Massive
  Star-forming Region W49A}},  {\em Astrophys. J.} {\bf 482} (1997), no.~1
  307--333.

\bibitem{Peng2010}
T.-C. Peng, F.~Wyrowski, F.~F.~S. van~der Tak, K.~M. Menten, and C.~M.
  Walmsley, {\it {W49A: a starburst triggered by expanding shells}},  {\em
  Astron. Astrophys.} {\bf 520} (Oct., 2010) A84.

\bibitem{Brun2011}
F.~Brun, M.~de~Naurois, W.~Hofmann, S.~Carrigan, A.~Djannati-Ata\"{\i}, S.~Ohm,
  and {for~the~H.~E.~S.~S.~Collaboration}, {\it {Discovery of VHE gamma-ray
  emission from the W49 region with H.E.S.S}},  {\em eprint arXiv:1104.5003}
  (2011).

\bibitem{Abramowski:2015rca}
{\bf HESS} Collaboration, A.~Abramowski et~al., {\it {The exceptionally
  powerful TeV gamma-ray emitters in the Large Magellanic Cloud}},  {\em
  Science} {\bf 347} (2015), no.~6220 406--412,
  [\href{http://arxiv.org/abs/1501.06578}{{\tt arXiv:1501.06578}}].

\bibitem{Mathewson1985}
D.~S. Mathewson, V.~L. Ford, I.~R. Tuohy, B.~Y. Mills, A.~J. Turtle, and D.~J.
  Helfand, {\it {Supernova remnants in the Magellanic Clouds. III}},  {\em
  Astrophys. J. Suppl. Ser.} {\bf 58} (June, 1985) 197.

\bibitem{Bamba2004}
A.~Bamba, M.~Ueno, H.~Nakajima, and K.~Koyama, {\it {Thermal and Nonthermal
  X‐Rays from the Large Magellanic Cloud Superbubble 30 Doradus C}},  {\em
  Astrophys. J.} {\bf 602} (Feb., 2004) 257--263.

\bibitem{Smith2004}
D.~A. Smith and Q.~D. Wang, {\it {Confronting the Superbubble Model with
  X‐Ray Observations of 30 Doradus C}},  {\em Astrophys. J.} {\bf 611} (Aug.,
  2004) 881--891.

\bibitem{Lacki2012}
B.~C. Lacki, {\it {Gamma-Ray Dominated Regions: Extending the Reach of Cosmic
  Ray Ionization in Starburst Environments}},  {\em eprint arXiv:1204.2580}
  (2012).

\bibitem{Romero:2003tj}
G.~E. Romero and D.~F. Torres, {\it {Signatures of hadronic cosmic rays in
  starbursts? High-energy photons and neutrinos from NGC 253}},  {\em
  Astrophys.J.} {\bf 586} (2003) L33--L36,
  [\href{http://arxiv.org/abs/astro-ph/0302149}{{\tt astro-ph/0302149}}].

\bibitem{Lacki:2010ue}
B.~C. Lacki and T.~A. Thompson, {\it {Diffuse Hard X-ray Emission in Starburst
  Galaxies as Synchrotron from Very High Energy Electrons}},  {\em
  Astrophys.J.} {\bf 762} (2013) 29,
  [\href{http://arxiv.org/abs/1010.3030}{{\tt arXiv:1010.3030}}].

\bibitem{Yoast-Hull2014}
T.~M. Yoast-Hull, J.~S. {Gallagher III}, E.~G. Zweibel, and J.~E. Everett, {\it
  {ACTIVE GALACTIC NUCLEI, NEUTRINOS, AND INTERACTING COSMIC RAYS IN NGC 253
  AND NGC 1068}},  {\em Astrophys. J.} {\bf 780} (Jan., 2014) 137.

\bibitem{Beacom2007}
J.~F. Beacom and M.~D. Kistler, {\it {Dissecting the Cygnus region with TeV
  gamma rays and neutrinos}},  {\em Phys. Rev. D} {\bf 75} (Apr., 2007) 083001,
  [\href{http://arxiv.org/abs/astro-ph/0701751}{{\tt astro-ph/0701751}}].

\bibitem{Gonzalez-Garcia2009}
M.~Gonzalez-Garcia, F.~Halzen, and S.~Mohapatra, {\it {Identifying Galactic
  PeVatrons with neutrinos}},  {\em Astropart. Phys.} {\bf 31} (July, 2009)
  437--444, [\href{http://arxiv.org/abs/0902.1176}{{\tt arXiv:0902.1176}}].

\bibitem{Fox2013}
D.~Fox, K.~Kashiyama, and P.~MŽszar—s, {\it {Sub-PeV Neutrinos from TeV
  Unidentified Sources in the Galaxy}},  {\em Astrophys.J.} {\bf 774} (2013)
  74, [\href{http://arxiv.org/abs/1305.6606}{{\tt arXiv:1305.6606}}].

\bibitem{Murase2015}
K.~Murase, D.~Guetta, and M.~Ahlers, {\it {Hidden Cosmic-Ray Accelerators as an
  Origin of TeV-PeV Cosmic Neutrinos}},  {\em eprint arXiv:1509.00805} (2015).

\bibitem{Torres2004}
D.~F. Torres, {\it {Theoretical Modeling of the Diffuse Emission of Gamma Rays
  from Extreme Regions of Star Formation: The Case of ARP 220}},  {\em
  Astrophys.J.} {\bf 617} (2004) 966--986,
  [\href{http://arxiv.org/abs/astro-ph/0407240}{{\tt astro-ph/0407240}}].

\bibitem{Yoast-Hull2015}
T.~M. Yoast-Hull, J.~S. {Gallagher III}, and E.~G. Zweibel, {\it {Cosmic Rays,
  Gamma-Rays, \& Neutrinos in the Starburst Nuclei of Arp 220}},  {\em eprint
  arXiv:1506.05133} (2015) [\href{http://arxiv.org/abs/1506.05133}{{\tt
  arXiv:1506.05133}}].

\bibitem{Hopkins:2006bw}
A.~M. Hopkins and J.~F. Beacom, {\it {On the normalisation of the cosmic star
  formation history}},  {\em Astrophys.J.} {\bf 651} (2006) 142--154,
  [\href{http://arxiv.org/abs/astro-ph/0601463}{{\tt astro-ph/0601463}}].

\bibitem{Ando2005}
S.~Ando, J.~F. Beacom, and H.~Y\"{u}ksel, {\it {Detection of Neutrinos from
  Supernovae in Nearby Galaxies}},  {\em Phys. Rev. Lett.} {\bf 95} (Oct.,
  2005) 171101, [\href{http://arxiv.org/abs/astro-ph/0503321}{{\tt
  astro-ph/0503321}}].

\bibitem{Kistler2011}
M.~D. Kistler, H.~Y\"{u}ksel, S.~Ando, J.~F. Beacom, and Y.~Suzuki, {\it
  {Core-collapse astrophysics with a five-megaton neutrino detector}},  {\em
  Phys. Rev. D} {\bf 83} (June, 2011) 123008,
  [\href{http://arxiv.org/abs/0810.1959}{{\tt arXiv:0810.1959}}].

\bibitem{Mattila2012}
S.~Mattila, T.~Dahlen, A.~Efstathiou, et~al., {\it {Core-collapse Supernovae
  Missed by Optical Surveys}},  {\em Astrophys J} {\bf 756} (2012), no.~111 15,
  [\href{http://arxiv.org/abs/1206.1314}{{\tt arXiv:1206.1314}}].

\bibitem{Botticella2012}
M.~T. Botticella, S.~J. Smartt, R.~C. Kennicutt, et~al., {\it {A comparison
  between star formation rate diagnostics and rate of core collapse supernovae
  within 11 Mpc}},  {\em Astronomy \& Astrophysics} {\bf 537} (2012) 21,
  [\href{http://arxiv.org/abs/1111.1692}{{\tt arXiv:1111.1692}}].

\bibitem{Odewahn1997}
S.~C. Odewahn, D.~Burstein, and R.~A. Windhorst, {\it {The Axis Ratio
  Distribution of Local and Distant Galaxies}},  {\em Astro.J.} {\bf 114}
  (1997) 2219, [\href{http://arxiv.org/abs/astro-ph/9709069}{{\tt
  astro-ph/9709069}}].

\bibitem{Cherry:2014xra}
J.~F. Cherry, A.~Friedland, and I.~M. Shoemaker, {\it {Neutrino Portal Dark
  Matter: From Dwarf Galaxies to IceCube}},
  \href{http://arxiv.org/abs/1411.1071}{{\tt arXiv:1411.1071}}.

\bibitem{Gross2010}
E.~Gross and O.~Vitells, {\it {Trial factors for the look elsewhere effect in
  high energy physics}},  {\em The European Physical Journal C} {\bf 70} (2010)
  525--530, [\href{http://arxiv.org/abs/1005.1891}{{\tt arXiv:1005.1891}}].

\bibitem{Choudalakis2011}
G.~Choudalakis, {\it {On hypothesis testing, trials factor, hypertests and the
  BumpHunter}},  \href{http://arxiv.org/abs/1101.0390}{{\tt arXiv:1101.0390}}.

\bibitem{botner2015}
O. Botner, talk at IPA2015 conference, Madison, WI, 2015. Available at:
  http://meetings.wipac.wisc.edu/IPA2015/home.

\bibitem{SAO1966}
S.~A.~O. Staff, {\it {Star catalog with positions and proper motions of 258,997
  stars for the Epoch and Equinox 1950.0}},  {\em Smithsonian Institution,
  Washington DC} (1966).

\bibitem{Han2014}
E.~Han, S.~X. Wang, J.~T. Wright, et~al., {\it {Exoplanet Orbit Database. II.
  Updates to Exoplanets.org}},  {\em Pub. Astro. Society of Pacific} {\bf 126}
  (2014) 827--837, [\href{http://arxiv.org/abs/1409.7709}{{\tt
  arXiv:1409.7709}}].

\bibitem{Blandford1979}
R.~D. Blandford and a.~Konigl, {\it {Relativistic jets as compact radio
  sources}},  {\em Astrophys. J.} {\bf 232} (Aug., 1979) 34.

\bibitem{Urry1995}
C.~M. Urry and P.~Padovani, {\it {Unified Schemes for Radio-Loud Active
  Galactic Nuclei}},  {\em Publ. Astron. Soc. Pacific} {\bf 107} (Sept., 1995)
  803.

\bibitem{Giommi:2011sz}
P.~Giommi, G.~Polenta, A.~Lahteenmaki, D.~Thompson, M.~Capalbi, et~al., {\it
  {Simultaneous Planck, Swift, and Fermi observations of X-ray and gamma-ray
  selected blazars}},  {\em Astron.Astrophys.} {\bf 541} (2012) A160,
  [\href{http://arxiv.org/abs/1108.1114}{{\tt arXiv:1108.1114}}].

\bibitem{Halzen1993}
F.~Halzen and R.~A. Vazquez, {\it {The GRO/Whipple Observation of Blazars:
  Implications for Neutrino Astronomy}},  {\em 23rd Int. Cosm. Ray Conf.} {\bf
  1} (1993).

\bibitem{Seyfert1943}
C.~K. Seyfert, {\it {Nuclear Emission in Spiral Nebulae.}},  {\em Astrophys.
  J.} {\bf 97} (Jan., 1943) 28.

\bibitem{Schmidt1983}
M.~Schmidt and R.~F. Green, {\it {Quasar evolution derived from the Palomar
  bright quasar survey and other complete quasar surveys}},  {\em Astrophys.
  J.} {\bf 269} (June, 1983) 352.

\bibitem{Weedman1977}
D.~W. Weedman, {\it {Seyfert Galaxies}},  {\em Annu. Rev. Astron. Astrophys.}
  {\bf 15} (Sept., 1977) 69--95.

\end{thebibliography}\endgroup

%%%%%%%%%%%%%%%%%%%%%%%%%%%%%%

\end{document}